\documentclass[aps, pra, twocolumn, notitlepage, nofootinbib]{revtex4-1}

\usepackage{graphicx,amsmath,pstricks,subfig}

\begin{document}

\title{Homogeneous cosmology with aggressively expanding civilizations}

\author{S. Jay Olson}
 \email{stephanolson@boisestate.edu}
 \address{Department of Physics, Boise State University, Boise, Idaho 83725, USA}

\date{\today}

\begin{abstract}
In the context of a homogeneous universe, we note that the appearance of aggressively expanding advanced life is geometrically similar to the process of nucleation and bubble growth in a first-order cosmological phase transition.  We exploit this similarity to describe the dynamics of life saturating the universe on a cosmic scale, adapting the phase transition model to incorporate probability distributions of expansion and resource consumption strategies.  Through a series of numerical solutions spanning several orders of magnitude in the input assumption parameters, the resulting cosmological model is used to address basic questions related to the intergalactic spreading of life, dealing with issues such as timescales, observability, competition between strategies, and first-mover advantage.  Finally, we examine physical effects on the universe itself, such as reheating and the backreaction on the evolution of the scale factor, if such life is able to control and convert a significant fraction of the available pressureless matter into radiation.  We conclude that the existence of life, if certain advanced technologies are practical, could have a significant influence on the future large-scale evolution of the universe.

\end{abstract}

\maketitle

\section{Introduction}

The vast majority of models describing the expansion of technological life into the cosmos have been designed at the scale of star-to-star travel, aimed at describing the process of colonizing the Milky Way~\cite{hart1975,jones1976,newman1981,bainbridge1984,landis1998,hanson1998} -- an approach that initially seemed most relevant for SETI.  Several developments now serve to highlight the possibility that the most relevant scale for the expansion of life could be cosmological, rather than a local neighborhood of stars -- these include a more detailed technological analysis of intergalactic and cosmological travel, showing that it is not significantly more difficult or expensive than interstellar travel~\cite{armstrong2013}, the continuing multi-decade null results from SETI searches within the Milky Way~\cite{tarter2001,shuch2011}, and recent proposals for and null results from the first ``intergalactic SETI" surveys~\cite{annis1999b,bradbury2011,carrigan2012,wright2014,wright2014b,calissendorff2013}.  This possibility raises the question of how a \emph{cosmological} process of life saturating the universe should be described, how to extract observable predictions from various starting assumptions, and what the physical consequences for the universe will be.

Here we describe such a process in the context of conservative assumptions on physics, but aggressive assumptions on technology.  That is, we assume known limitations imposed by fundamental physics will not be overturned by new discoveries made by advanced civilizations, but we assume several ambitious ``in principle" technological concepts will become practical for maximally advanced civilizations.  In particular, we base our analysis on the following core assumptions:
\begin{enumerate}
\item At early times (relative to the appearance of life), the universe is described by the standard cosmology -- a benchmark Friedmann-Robertson-Walker (FRW) solution\footnote{We use a spatially flat FRW solution with $ \Omega_{\Lambda 0} = .683$, $\Omega_{r 0} = 3 \times 10^{-5} $, $\Omega_{m 0} = 1-\Omega_{\Lambda 0} - \Omega_{r 0} $, and $H_0 = .069 \ Gyr^{-1} $, putting the present age of the universe at $\approx 13.75$ Gyr. We work in co-moving coordinates and use Gyr and Gly as our units of choice.  The scale factor is denoted by $a(t)$.}.
\item The limits of technology will allow for self-reproducing spacecraft, sustained relativistic travel over cosmological distances, and an efficient process to convert baryonic matter into radiation.
\item Control of resources in the universe will tend to be dominated by civilizations that adopt a strategy of aggressive expansion (defined as a frontier which expands at a large fraction of the speed of the individual spacecraft involved), rather than those expanding diffusively due to the conventional pressures of population dynamics. 
\item The appearance of aggressively expanding life in the universe is a spatially random event and occurs at some specified, model-dependent rate.
\item Aggressive expanders will tend to expand in all directions unless constrained by the presence of other civilizations, will attempt to gain control of as much matter as is locally available for their use, and once established in a region of space, will consume mass as an energy source (converting it to radiation) at some specified, model-dependent rate.
\end{enumerate}

These assumptions suggest a physical role for life which is substantially different than usually supposed.  Instead of life as a barely-noticeable feature surviving on top of a completely indifferent background cosmology, we imply that the behavior of advanced life could ultimately become a serious variable in the large-scale physical description of the universe, making substantial alterations to the matter composition and entropy on a relatively short timescale, while inducing a small but measurable backreaction on the evolution of the cosmic scale factor and Hubble parameter.  

Describing such possibilities in detail must inevitably take the form of a cosmological model, and our goal will be to construct a framework for this kind of model-building, to examine the issue of observability, and to numerically evaluate a selection of such models to explore features of interest.  Owing to the homogeneity of a FRW universe, our core assumptions allow for geometrical simplicity in describing the aggressive expansion of advanced life.  The various underlying processes (such as the rate of production of habitable planets, probabilities of biological evolution, technology and behavior patterns, etc.) are encapsulated into a few basic spatially averaged functions and parameters, which may be modeled to any desired degree of complexity.  Such an approach encourages the examination of many scenarios with parameters varying over many orders of magnitude, and to explore the maximum possible influence of life on the universe itself, by coupling the Friedmann equations to the waste heat of a universe saturating with maximally-advanced life.

Our expansion model will be built from the following sets of functions and parameters, including the \textit{appearance rate} $f_i(t)$ of aggressively expanding life per unit coordinate volume (co-moving coordinates) per unit time, the \textit{saturation volume} $V_i(t',t)$ at time $t$ for a species that appeared at time $t'$, and the \textit{saturation time} $T_i$, which describes the length of time between the initial arrival of the first spacecraft at the frontier, and a visible change to the matter there (due to the assumed aggressive use of resources).  The index $i$ runs over all behavior types considered.  This type of model, in the case of only one behavior type and $T=0$, corresponds exactly to a model of nucleation and bubble growth for a first-order cosmological phase transition\footnote{The analogy with a phase transition has been made before, in the context of ET life, to describe a scenario in which the evolutionary progression of life throughout a single galaxy becomes correlated due to galaxywide extinction events~\cite{annis1999}.  The general idea has since been referred to as a ``global regulation mechanism,"~\cite{vukotic2008} but in any case it is distinct from what we are doing here.  }.  An illustration of the basic picture appears in figure 1.

\begin{figure}
\subfloat[]{
  \includegraphics[width=0.45\linewidth]{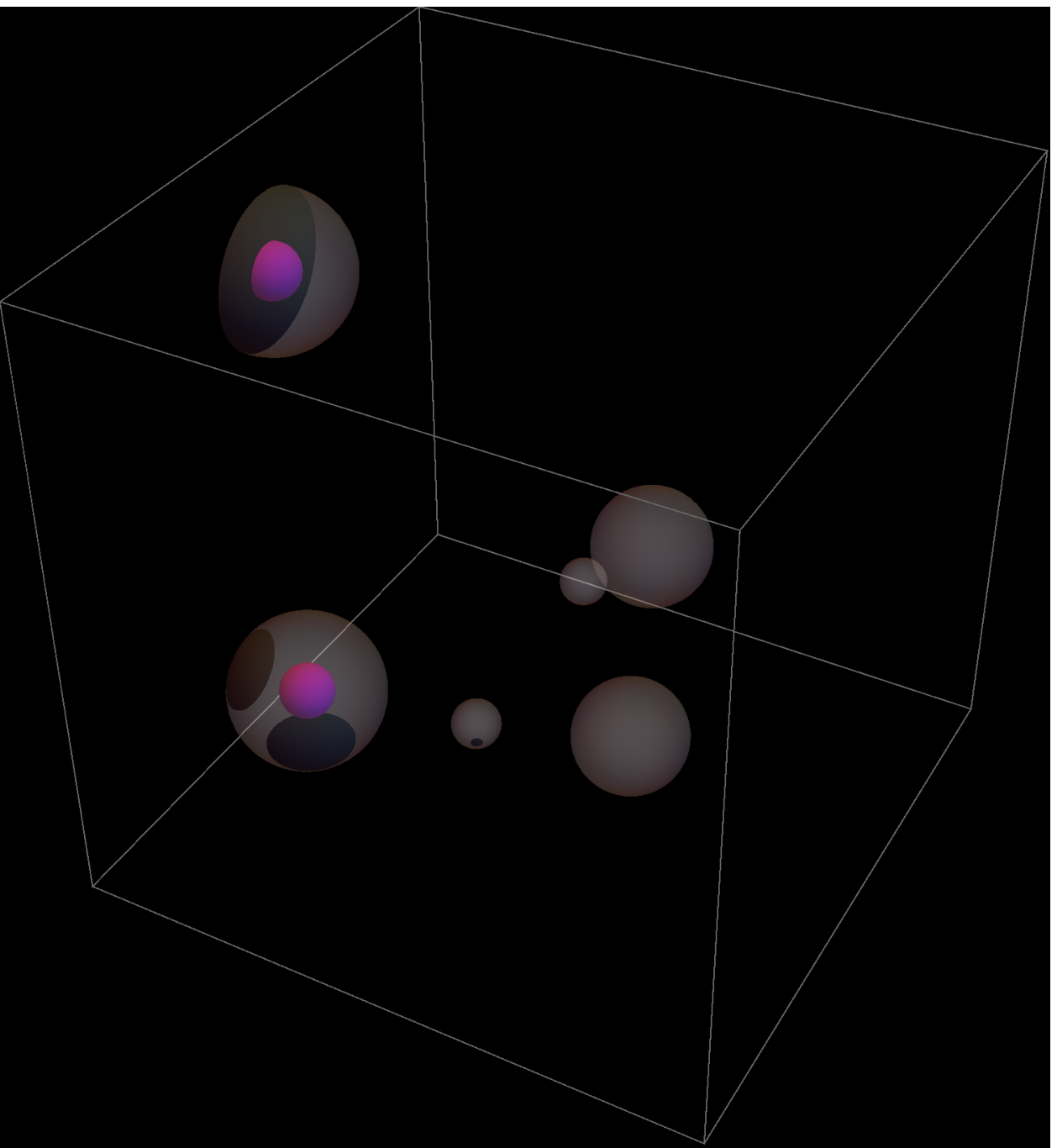}
}
\subfloat[]{
  \includegraphics[width=0.45\linewidth]{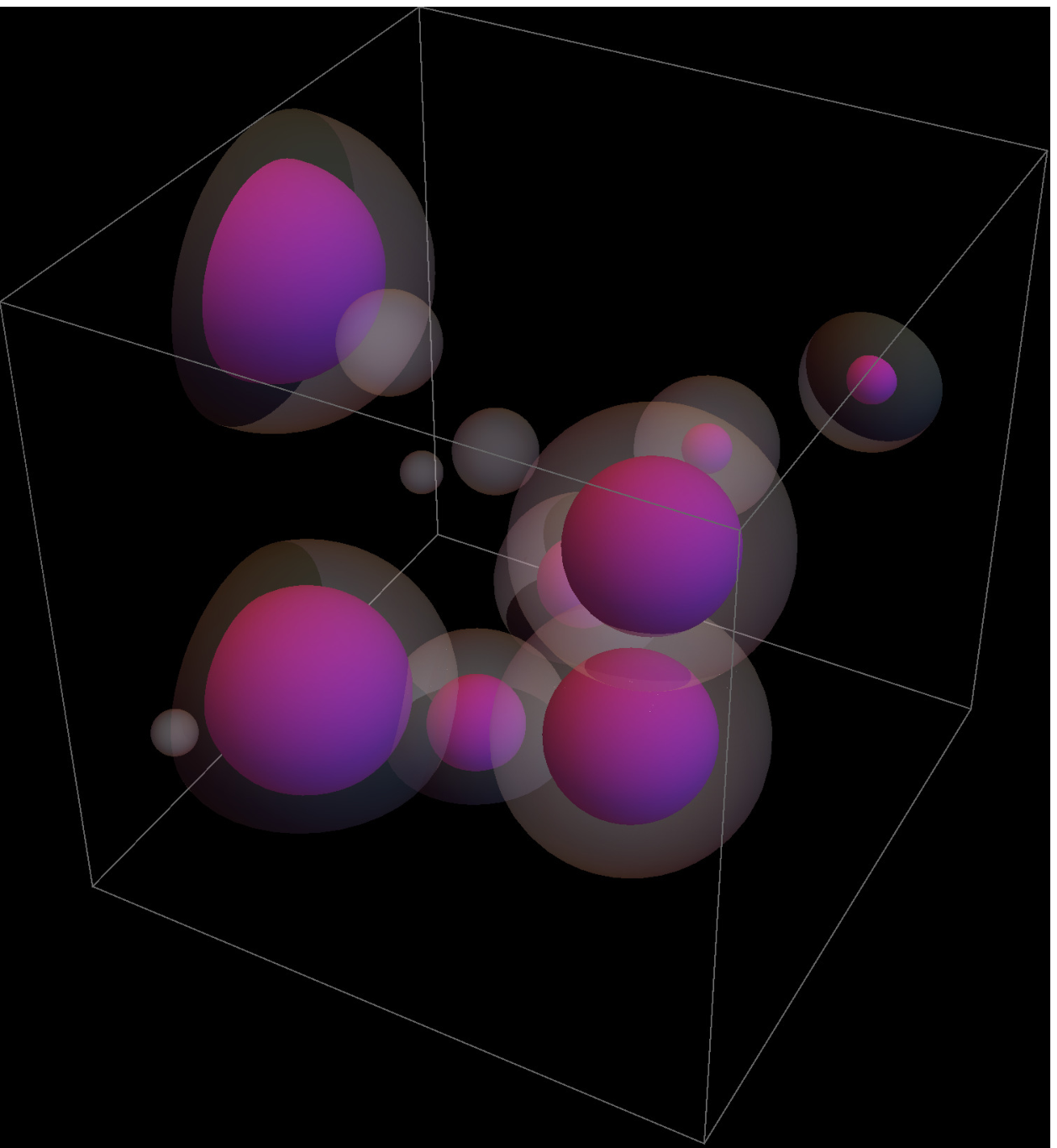}
}
\hspace{0mm}
\subfloat[]{
  \includegraphics[width=0.45\linewidth]{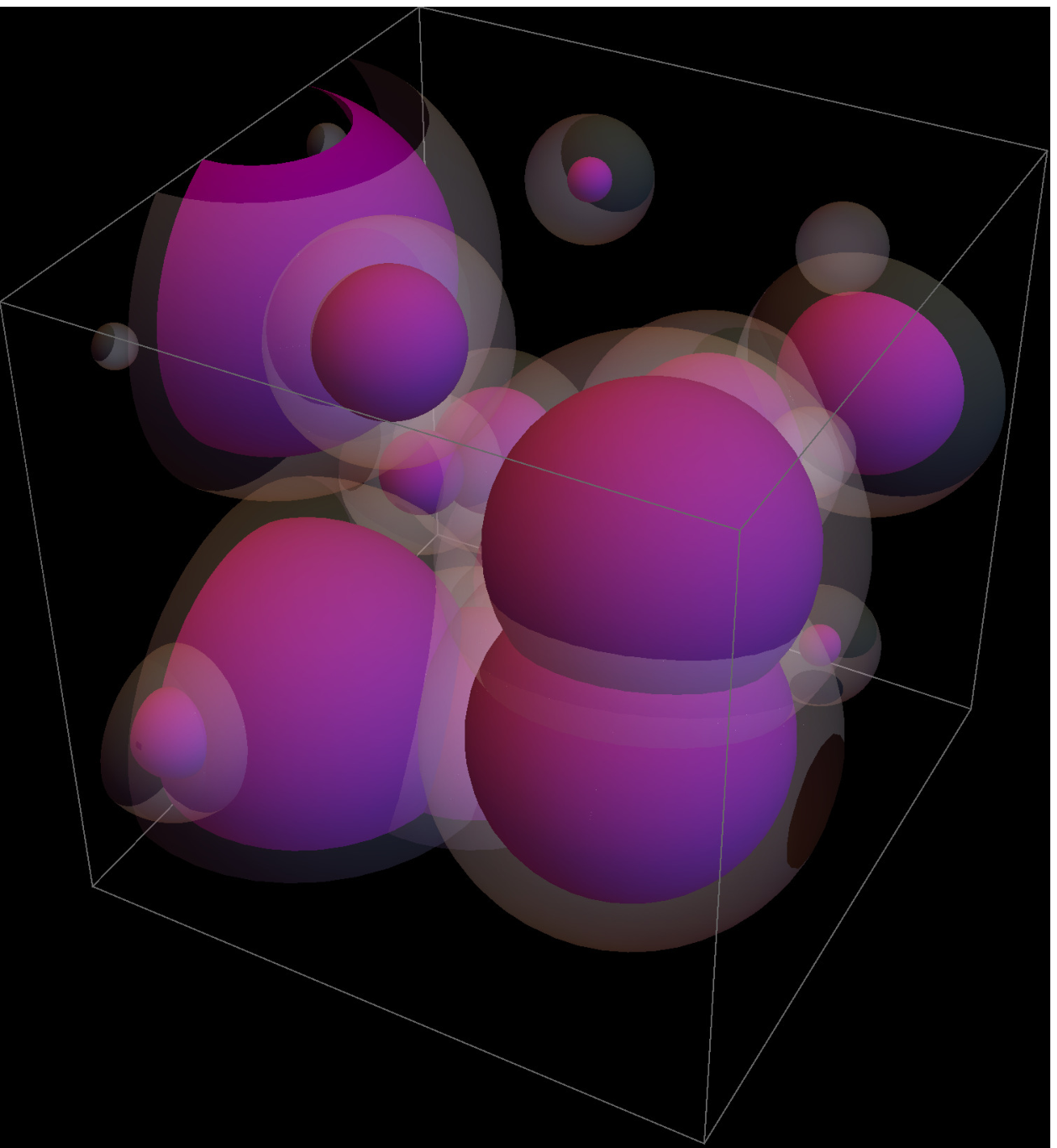}
}
\subfloat[]{
  \includegraphics[width=0.45\linewidth]{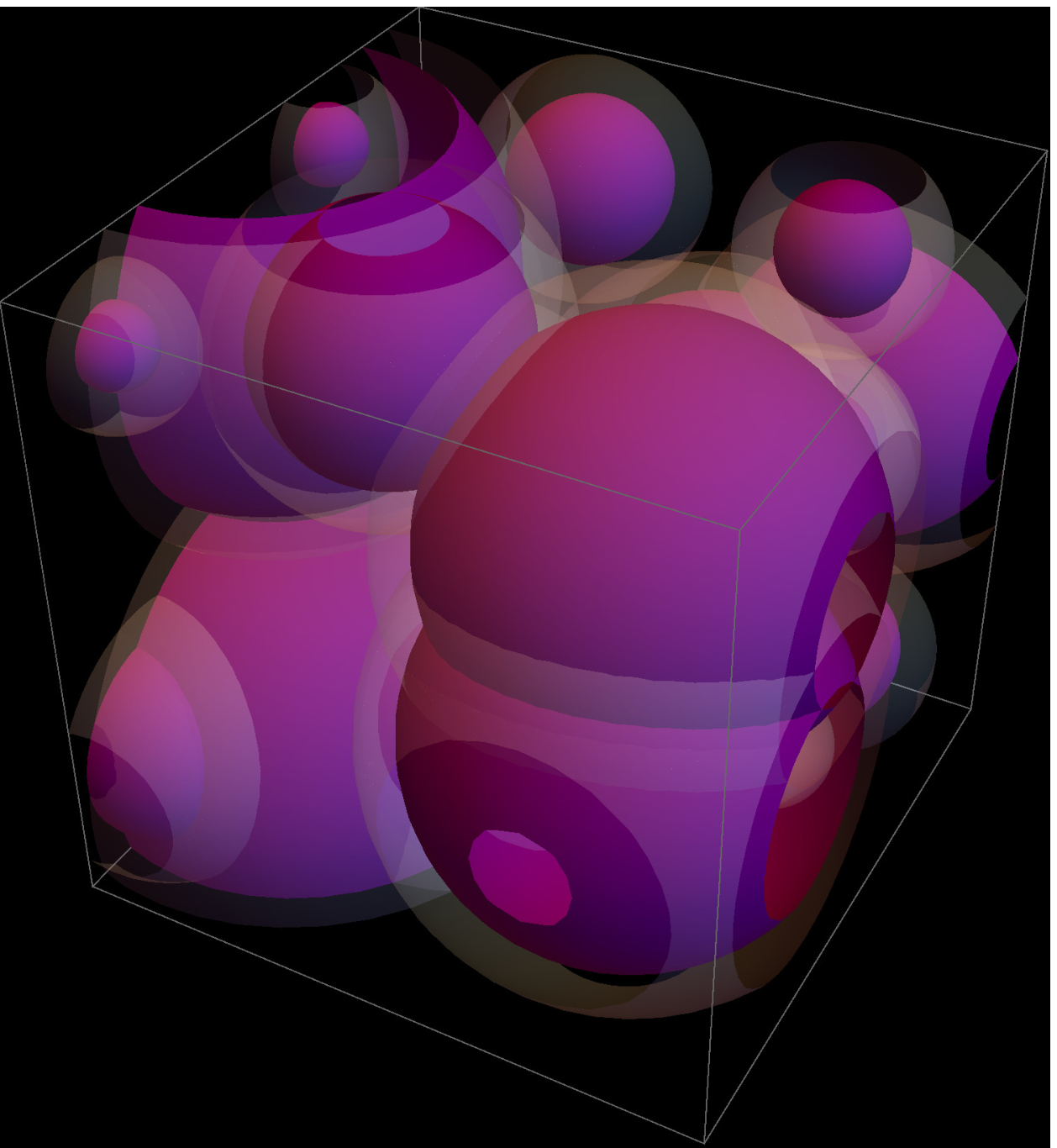}
}
\caption{An illustration of the time-development of life saturating a rectangular portion of the universe.  The transparent spheres represent the initial diffuse front of spacecraft released by advanced civilizations appearing randomly throughout the universe.  The opaque spheres represent the front of fully life-saturated matter (appearing a time $T$ after the probe front has passed by), whose waste heat or large scale engineering efforts are presumed to be visible across cosmic distances.  (a)-(d) are snapshots at equal time intervals.  The geometry and dynamics of the universe filling with saturated matter in this way is analogous to bubble formation and growth in a first-order cosmological phase transition. }
\end{figure}

Because the assumption of homogeneity becomes valid at a distance scale of approximately 0.25 Gly, this approach will be most appropriate in the regime where the appearance rate for aggressively expanding strategies is very small, allowing early civilizations to expand over distances larger than the homogeneity scale before being constrained by other expanding civilizations.  Such a requirement is consistent with both of the two most obvious interpretations of the ``great silence~\cite{brin1983}," where either the evolution of advanced technological life is much rarer than typically supposed~\cite{tipler1980}, or else the vast majority of such life chooses not to expand~\cite{sagan1983}.  In the latter case, non-expanding life will saturate an insignificant fraction of the universe's resources, and will not constrain the growth of rapidly expanding civilizations.  There is a certain irony in noting that the greater the probability of advanced life to become inward-focused and to not expand, the greater will be the applicability of a cosmological model for describing the relatively rare but rapidly expanding civilizations.

We organize our work in the following way:  Because \emph{any} input assumptions on this topic are inevitably subject to controversy, we begin in section 2 with a discussion and fleshing-out of our core assumptions regarding the appearance, motives, and behavior of advanced life, and the limits of technology.  Some of these elements will be critical to the framework, while others can be modified as needed, corresponding to adjusting a parameter or other relatively simple changes to the mathematical framework.

Due to the geometrical similarity to a phase transition, section 3 will be a review of two of the most common descriptions of nucleation and bubble growth in a first-order cosmological phase transition, which give dynamical expressions for the transformed fraction of space.  Section 4 is the theoretical construction of the framework, where we generalize the phase transition models to include a distribution over behavior types, and couple the possible effects of life to the Friedmann equations.  We then obtain expressions for determining relative strategy success, first-mover advantage, and the observability of the process from a generic vantage point.  By introducing a non-zero saturation time $T$, we are implying that the rapidly expanding but diffuse ``probe front" is not observable, but the ``saturation front" (of large-scale engineering efforts -- Dyson spheres~\cite{dyson1960}, stellar deconstruction, waste heat~\cite{wright2014}, etc.) following after time $T$ should be observable over great distance.  An interesting exception takes place if the probe front is sufficiently fast and the saturation time is sufficiently long;  the expansion of the universe then implies that the saturation front advances with superluminal speed\footnote{There is no conceptual tension between causality and the superluminal propagation of the saturation front -- it is the probe front (propagating subluminally) that carries the information required to generate the saturation front, which is constructed locally after time $T$.}, and it becomes effectively impossible to observe from a distance the expansion of life that uses such a strategy.  Although such a scenario requires parameters that may seem extreme, it is the limiting case of a more general phenomenon -- the faster the expansion and the longer the saturation time, the more narrow the window to observe cosmic saturation in progress before it overtakes one's vantage point.

In section 5, we examine numerical solutions for a collection of scenarios that illustrate some general features of interest.  Among these are the limited observability of the fastest strategies, competition between fast vs. common strategies, reheating of the universe, and the cosmic age-dependence of relative strategy success.  

The final sections are devoted to additional conclusions, and our acknowledgments.

\section{Underlying assumptions}
The ingredients of our analysis are conventional physics in a very broad sense, though in keeping with the theme of \emph{aggressive} expansion, we will deliberately push the limits of what may be practically achievable for a maximally advanced civilization.  In any case, however, we exclude the possibility of faster-than-light travel or violations of the dominant energy condition.  The description utilizes a solution of the standard cosmology.  Though always subject to some debate, we will not explore possible consequences of modifying these ingredients here. Far greater uncertainties arise regarding the appearance, motives, and behavior of advanced life, and the limits of technology, and so for this reason we discuss in more detail those assumptions here.

\subsection{Technology}
First, we assume that for any relevant species, the development of maximally-advanced technology is a sudden event on the timescales considered.  This follows the thinking of various authors who have described the concept of an ``intelligence explosion,"~\cite{good1965} or a ``technological singularity,"~\cite{vinge1993,kurzweil2005} where increasingly great technological advances occur in an increasingly compressed period of time\footnote{Possible modifications of the basic idea, such as replacing exponential technology growth with a logistical function, should not change the basic conclusion for our purposes.}.  As a consequence, we assume that full mastery of the laws of physics occurs in a short time interval, so that all expanding species considered have already reached a point of maximum technological sophistication.  For our purposes, we will take this to mean that expansion and resource consumption strategies will not evolve due to continuing technological advancements after the initial release of aggressively expanding spacecraft, and that once a species of ``behavior type $i$" has saturated the matter in some region of space, we will not need to consider the possibility of another, far more advanced species of ``behavior type $j$" re-taking it.  Other technological trajectories are possible, e.g. abandoning the development of powerful but dangerous nano-scale technology in an attempt to minimize the risk of planetary disasters~\cite{joy2000}, but civilizations choosing such a path cannot aggressively expand in the manner we will describe, and will end up occupying an insignificant fraction of the universe, effectively removing themselves from the analysis.

We will assume that relativistic travel is practical over cosmological distances.  This possibility has been considered in more detail elsewhere, usually in the context of a spacecraft traveling along a geodesic trajectory~\cite{tipler1999,armstrong2013}.  We will consider, however, a class of space probes which have the capability of sustaining some maximum practical velocity relative to the local co-moving frame (implying a non-zero proper-acceleration in the reference frame of the individual probes) -- this would seemingly not be a difficult requirement for a spacecraft powered by a black-hole drive of the type described by Crane and Westmoreland~\cite{crane2009}, if combined with a magnetic ram scoop to harvest fuel either in the vicinity of individual galaxies or in the intergalactic medium itself~\cite{bussard1960}.  We also allow for self-reproduction on the part of the probes, using collected interstellar and/or intergalactic media.  In the most aggressive technological scenario, this would require the probes to fuse their own heavy elements from the light elements they collect over the vast timescales of their travel (though a periodic ``stop to collect resources and reproduce" strategy would also qualify as aggressive expansion, so long as the stopping time is sufficiently small compared to the intergalactic travel time).  Cosmological travel and self-reproducing capability are really essential assumptions for this model, enabling the behavior assumptions we describe below.  If it turns out that fundamental limitations to technology prohibit spacecraft with these capabilities, the behavior assumptions and expansion model will need to be heavily modified. 

The least conservative technological capability we consider is that maximally advanced life will be able to convert a significant fraction of the pressureless matter of the universe into waste radiation, over a relatively short period of time (compared with other natural processes), as an energy source for their civilizations.  The pressureless matter of the universe (i.e. the major source of energy density that is not radiation or dark energy) is often referred to as ``dust" and is composed primarily of dark matter and baryons.  The dark matter component is commonly believed to be weakly interacting massive particles, and we assume this implies that the vast majority of dark matter will remain permanently outside the control of advanced civilizations.  A large majority of the baryonic component of the dust exists in the form of ionized gas, either held by galaxies or in the intergalactic medium -- stars themselves account for less than 10\% of the baryonic matter in the universe~\cite{prochaska2009}.  Our assumption is that over the course of many billions of years, a significant fraction of these baryons will eventually be harvested in a universe that becomes filled with maximally advanced civilizations.  The fraction of baryons remaining permanently inaccessible to such civilizations will thus represent a kind of fundamental limitation imposed by physics, and we assume that it will be a uniform limitation for all aggressively expanding civilizations.  The scenarios we will describe involve the conversion of $1 \% - 5 \%$ of the cosmic dust into radiation.

The conversion process itself could presumably be accomplished via this large-scale collection of massive particles, feeding them into microscopic black holes which convert the mass into blackbody radiation at the Hawking temperature, which in turn is converted into useful work with heat engines~\cite{crane2009}.  Of course, we are in no position to say that this represents the most efficient or feasible technology for converting mass into work (and eventually to waste radiation) -- it is used only as an illustration of principle that the laws of physics seem to allow such processes.  Indeed, Hawking radiation is presently understood only at the semiclassical level, suggesting that major refinements to the basic concept may eventually become available, while the limiting thermodynamics of \textit{feeding} a microscopic black hole have not yet been described, nor have practical limits to collecting mass in the universe been explored.  On the one hand, this means that our estimates for the depth and speed of life consuming mass as fuel are effectively pure guesses at this stage.  On the other hand, it breaks nothing in the model if our estimates turn out to be unreasonable -- modifying this assumptions is as simple as adjusting a parameter, which can be taken to zero if necessary. 

\subsection{Motives}
We assume that the motives for aggressive expansion are one possible continuation of the motives which lead to maximizing technological ability.  Once fundamental limits to technology are reached on a home world, the only way to continue to increase capabilities will be to expand and coordinate the use of more resources~\cite{kurzweil2005}.  Others have argued that advanced life will tend to become inward focused and avoid expansion~\cite{cirkovic2008} -- we do not contest this possibility, provided that \textit{some} fraction of advanced species adopt aggressive expansion as a means of increasing their capabilities (in fact, consistency of our model with observation requires that aggressive expansion is a rare event in the universe). 

In studying aggressive expansion, then, we are \emph{not} examining a traditional process of slow biological diffusion driven by the usual pressures of population dynamics.  Instead, we are examining a process more akin to a cosmic-scale engineering effort with some pre-determined goal that requires coordinating the use of vast quantities of matter and energy.  What sort of goal this could be remains speculation at this time -- construction of a Gly-scale brain, or engineering the matter distribution to create a locally static universe (to dramatically slow the causally isolating effect of an accelerating universe~\cite{krauss2007}) might be two such possibilities.  We regard the possibility that advanced intelligence will make use of the universe's resources to simply populate existing earthlike planets with advanced versions of humans as an unlikely endpoint to the progression of technology.

We ascribe importance to this aggressive sort of expansion because our technology assumptions make it simultaneously dominant (over slow, diffusive migration) and cheap (the cost is exactly one probe with general replication capabilities and sufficiently detailed instructions).  It could still be the case that most advanced civilizations choose not to aggressively expand, but the barrier appears to be very low to doing so, and those making the decision to do so will end up controlling essentially all of the resources of the universe.

Debates on the use of self-replicating spacecraft (von Neumann probes~\cite{freitas1980}) to explore the galaxy, originating in the 1980's~\cite{tipler1980,sagan1983}, were well-aware of the powerful implications of this kind of technology.  The original objection~\cite{sagan1983} to the presumed use of such technology in the galaxy, however, seemed to focus on a type of space probe with surprisingly primitive instructions, such that it would simply reproduce itself forever, consuming all resources in a manner counter-productive to the purposes of its original designers.  Our position is that this sort of reasoning was never really an argument against the \emph{general} use of self-reproducing technology in the universe, but rather an argument against a \emph{special case} of instructions (or instructions sufficiently vulnerable to random mutation -- a technological consideration that seems to present no fundamental barrier~\cite{wiley2011}) which would generally not be useful.     

Another motivating factor for expansion and rapid use of resources comes from the ultimate destiny of the universe -- the expansion of the universe appears to be ever-accelerating, with the consequence that the amount of accessible matter declines rapidly as it falls across the cosmological event horizon.  Eventually, all that remains are causally isolated galaxy clusters~\cite{krauss2007}.  Thus, if a civilization desires any coordination of resources beyond this scale, it is advantageous to move as quickly and aggressively as possible. 

Finally, aside from rational incentives to expand, it is also possible that an accidental AI catastrophe results in aggressive expansion -- in this scenario, a recursively self-improving AI with a primitive utility function (e.g. ``maximize performance") is switched on without sufficient safeguards, and the resulting intelligence finds no reason to stop at the boundaries of its home world, solar system, or galaxy.

\subsection{Behavior}

Based on the above assumptions, we imagine the following aggressive expansion scenario:  A recently evolved advanced species releases a wave of probes (``expanders") in all directions.  They are designed to travel at some maximum velocity $v$ in the local co-moving frame of reference, and they scoop material from the interstellar and/or intergalactic medium in a manner analogous to the Bussard ramjet~\cite{bussard1960}.  As the expanders travel over vast distances and times, they are designed to reproduce themselves and adjust their velocity slightly at pre-determined intervals, so that the expanding sphere of probes maintains a roughly constant density.  Also at pre-determined intervals, the expanders are designed to construct a different type of probe (a ``seed probe") whose job it will be to decelerate to the local co-moving frame of reference, reproduce itself from the available matter, and saturate some specified volume of space for whatever purposes the probes were initially released -- a process that could take a very long time (which we associate with the parameter $T$).  Once matter has been saturated, life will begin to ``burn" mass by converting it into radiation.  The fraction of the cosmic dust available for fuel is $A$ (the inaccessible fraction is $I = 1 -A$), and we assume that the rate of fuel consumption will be proportional to the amount of fuel saturated, resulting in an effectively exponential decay law for the saturated matter, which we specify with decay constants $\Gamma_i$.

We do not require the details of such a strategy to be universal amongst expanding species, but the existence of such a strategy, consistent with our technology assumptions, motivates our interest in studying highly aggressive expansion with a range of parameters that include relativistic $v$, and we will find in section 5 that high-$v$ strategies tend to dominate the universe, even if their appearance rate is small compared to low-$v$ strategies.  The ``exponential decay" resource consumption model is chosen mainly for its intuitive simplicity, and incorporating a more general model of resource consumption is a likely future improvement of the framework. 

We have not specified the detailed process of reproduction and spreading of the decelerating seed probes which ultimately perform the work of saturating a large volume of available matter.  We have only assumed that a constant density of expanders on the probe front implies a constant time delay $T$ for the saturation process (since in this case every released seed probe has an identical volume of space to saturate).  A more detailed model of the saturation process will allow for a more general, time dependent saturation time $T(t,t')$.  We regard this as a next-order correction to the framework, as in addition to the more specific assumptions required, such modeling will tend to involve numerical solutions to delay differential equations with general time-dependent delays.  For now, we simply assume that our constant $T$ represents an approximate final value of the saturation time after the density of the probe front has come to some optimal value specified in the original probe instructions.

It is also possible and perhaps quite reasonable to expect that many simultaneous ``species" of expanders would be released from the same civilization with different programming and competitive goals -- we expect that the net effect of such a scenario would be highly aggressive expansion and resource use~\cite{hanson1998}, and should still be described by a probe front velocity, a characteristic time to make significant changes to matter, and a resource consumption function, though with more detailed models for $T$ and $\Gamma$ than we introduce here. 

\subsection{Appearance}
The appearance rate of advanced life has been highly controversial for decades, but in the present context we are only interested in the appearance rate for the aggressively expanding subset of civilizations.  We assume these will appear randomly throughout the universe and in proportion to the number of earth-like planets orbiting main-sequence stars, which have had sufficient time for intelligence to evolve.  For purposes of simple modeling, we assume that it takes at least 4.5 Gyr for intelligence to evolve on a newly formed earth-like planet, and that the window for life to evolve (constrained by stellar evolution) is 6 Gyr \footnote{This window implicitly ignores the possibility of advanced life appearing around stable M-dwarfs, for which the habitable zone lifetime would be dramatically longer.}.    

With these assumptions in mind, we make use of a simplified version of Lineweaver's model for the formation rate of earth-like planets in the universe~\cite{lineweaver2001}.  To accomplish this, we use a star formation rate $SFR(t)$ that approximates the basic features of Lineweaver's fit, namely that $SFR(0)=0$, and that $SFR$ increases exponentially until a maximum at $t=3$, where it then begins an exponential decline until it is reduced by one order of magnitude at present day ($t=13.75$).  The simple function (normalized to unity at $t=3$) we use to accomplish this is given by:
\begin{displaymath}
   SFR(t) = \left\{
     \begin{array}{lcc}
       \frac{t}{3} 10^{t-3} & : &   t < 3  \\
       10^{-\frac{(t-3)}{13.75 - 3}} & : &  t \geq 3 
     \end{array}
   \right.
\end{displaymath} 
We now integrate the star formation rate to give a function that will be proportional to the buildup of metallicity in the universe, i.e. $M(t) = \int_{0}^{t} SFR(t') \, dt'$.  We now use a simplified approximation to model the formation rate of earth-like planets as $PFR(t) = M(t) \, SFR(t) $, again normalizing $PFR$ to unity at its maximum value.  The assumption that the appearance rate of aggressively expanding life is proportional to earthlike planets that are older than 4.5Gyr (time enough for evolution) but no older than 6Gyr (an approximate lifetime of a habitable zone) then gives our appearance function:
\begin{displaymath}
f(t) = \alpha \int_{t-6}^{t-4.5} M(t') \, SRF(t') \, dt'.
\end{displaymath}
The parameter $\alpha$ sets the basic scale for the appearance rate of expanding life, and is modulated by an integral which is presently of order unity, but is exponentially suppressed before $t \approx 7.5$Gyr (and is exactly zero before $t=4.5$Gyr).  This model is simplified in the sense that we do not model a distribution over individual star metallicities, and so do not take into account the likely destruction of earthlike planets around extremely high-metallicity stars, or the promotion of planet formation when the average metallicity of the universe is still very low.  One can nevertheless see from figure 2 that our planet formation rate approximates the basic features of the Lineweaver model quite well.

In what follows, we will vary our appearance rate model through the single constant, $\alpha$.  We emphasize that simple changes to this model (such as postponing the evolution of life due to hostile early conditions) can result in very substantial changes to the solutions we obtain, as well as the implied current presence and observability of life in the universe, though we will not delve into such alternative models here.

The appearance rate parameters $\alpha$ we will study extend over five orders of magnitude, from $\frac{1}{100 \, Gly^3 \,Gyr }$ to $\frac{100,000}{100 \, Gly^3 \,Gyr }$.  The entire range is much smaller than traditionally supposed (though again we emphasize that we consider one subset of total advanced life), with the upper end of the range corresponding roughly to one aggressively expanding species per galaxy supercluster per billion years.  This range has been chosen purely to generate convenient timescales for saturation of the universe when the probe front is a significant fraction of the speed of light ($ \geq .01 c$).  Nonrelativistic expansion (or the aforementioned possible delay in the appearance model) allows much higher appearance rates to be consistent with a neighborhood that seems devoid of aggressively expanding life, but a rate very much higher would presumably become an obvious feature of ordinary galaxy surveys.

\begin{figure}
\subfloat[]{
  \includegraphics[width=1\linewidth]{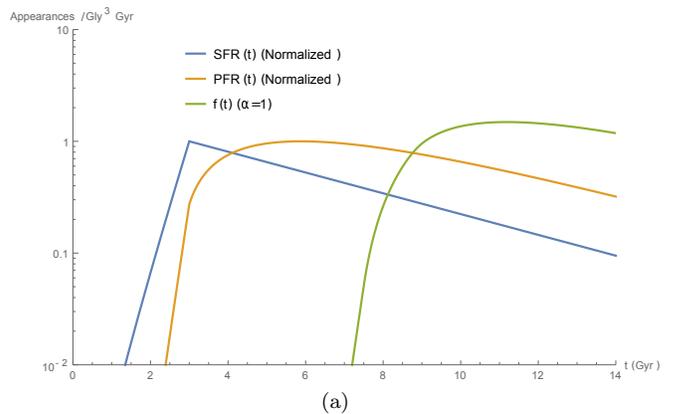}
}

\caption{Star formation rate $SFR(t)$, planet formation rate $PFR(t)$, and the appearance rate for aggressively expanding life $f(t)$ (for $\alpha=1)$, which becomes the nucleation rate for our model.  $SFR$ and $PFR$ are normalized to a maximum value of unity, while $f(t)$ has units of appearances per $\mathrm{Gly}^3$ of co-moving coordinate volume per $\mathrm{Gyr}$ of cosmic time.  }
\end{figure}

\section{Review of Nucleation and bubble growth in first-order cosmological phase transitions}

Models of nucleation and bubble growth in a phase transition are the geometrical basis for our analysis, and so a brief digression should be useful for those with a different background.

In a first-order phase transition, the existence of a potential barrier prevents a spatially uniform transition from a metastable vacuum to the true vacuum as temperature is lowered.  Instead, thermal fluctuations probabilistically overcome the potential barrier as the temperature falls below criticality.  When such a random fluctuation takes place, a bubble of the new phase is nucleated.  If the bubble is initially of sufficient size, it will then grow at high speed.  Eventually, enough bubbles have formed and grown so that the entire universe is in the new phase (see figure 1).

To calculate the fraction of space remaining in the old phase, $g(t)$, we would need the nucleation rate per unit time per unit coordinate volume, $f(t)$ (derived from a quantum mechanical calculation), and we need the coordinate volume of space occupied by a single bubble nucleated at time $t'$, $V(t',t)$.  In a FRW universe, the latter is given by $V(t',t) =\frac{4 \pi}{3} (\int_{t'}^{t} \frac{v \theta(t'' - t')}{a(t'')} dt'')^3 $ (where $\theta$ is the Heaviside step function), assuming that the bubbles expand with velocity $v$ in the comoving frame.  Next, we need to specify what happens when expanding bubbles collide and merge -- different assumptions result in substantially different formulas for $g(t)$.

Two of the most popular approaches are given by Guth-Tye-Weinberg (GTW)~\cite{guth1980,guth1981}, and Csernai-Kapusta (CK)~\cite{csernai1992}.  In the GTW approach, the bubble walls pass through one another and continue expanding from their respective centers -- i.e. the collision does not effect the spherical geometry of individual bubbles.  If we assume that ``virtual bubbles" continue to nucleate and grow within real bubbles (being inside a virtual bubble presumes already and forever being within a real bubble, so $g(t)$ is unaffected by this assumption), this allows us to regard the probability of any point in space being within bubbles nucleated at different times as independent events, so that we can express the probability of not being in any bubble nucleated between times $t_i$ and $t_f$ as $g(t_i,t_f ; t) = g(t_i,t_m; t)g(t_m,t_f ; t)$, where $t_i < t_m < t_f$.  This enables the following:
\begin{eqnarray}
g(0,t'+dt' ;t) = g(0,t';t)g(t',t' + dt';t)\\
 = g(0,t';t)[1 - f(t')V(t',t)dt']\\
 \frac{d g(0,t';t)}{dt'} = -g(0,t';t)f(t')V(t',t).
\end{eqnarray}
The latter equation may be integrated to obtain $g(t)$,
\begin{eqnarray}
g(t) \equiv g(0,t;t) = e^{- \int_{0}^{t} f(t') V(t',t) dt' }
\end{eqnarray}
which is the Guth-Tye-Weinberg formula.

It was later argued that the assumption of perfectly overlapping spheres underestimates the growth of colliding bubbles, and an approximation was proposed in which the volume of two colliding bubbles continues to increase as the sum of their individual volumes.  In this approach, one does not invoke virtual bubbles, and instead asserts that the rate of formation of new bubbles is proportional to the nucleation rate times the fraction of space remaining in the old phase, i.e. $f(t) g(t)$.  Since each bubble expands with volume function $V(t',t)$, unaffected by collisions with other bubbles, we are led to the following integral equation:
\begin{eqnarray}
g(t) = 1 - \int_{0}^{t} f(t')g(t')V(t',t) \ dt'
\end{eqnarray}
which is the Csernai-Kapusta formula, which tends to overestimate the fraction of space in the new phase whereas the GTW formula underestimated it.  In terms of probability, the key assumption of GTW is that of independent events, while the key assumption of CK is that of mutual exclusivity. 

In our context, where bubbles represent the expansion of probes, and a jump in stress-energy at the bubble wall is not the significant factor driving expansion (but is instead the initial design of the probes), we assume that ``overlapping spheres" gives the correct geometry, favoring the GTW approach.  However, in the general case, with a distribution of bubbles expanding at various velocities, we can no longer invoke the GTW formula or the reasoning of virtual bubbles and independent events -- this is due to the possibility of fast virtual bubbles breaking out of slow real bubbles, leading to an overestimate of the fraction of space saturated by fast strategies.  Instead, we note that the CK formula agrees with the GTW formula if the volume function $V(t',t)$ appearing in the CK equation is changed to reflect the assumption of overlapping spheres, via $\bar{V}(t',t) = \frac{1}{g(t')} \int_{t'}^{t} dt'' g(t'') \frac{d V(t',t'')}{dt''}$ -- that is, we integrate the growth rate of a sphere times the fraction of unsaturated space it expands into, and average only over ``real" spheres formed at $t'$.  This allows us to simultaneously use the property of mutual exclusivity and the geometry of overlapping spheres, and will be the basis for generalizing the phase transition framework to a distribution of velocities $v_{i}$ in the next section.

It is also important to note the extent of homogeneity in our analysis of aggressive expansion.  Strictly speaking, the process we will describe is homogeneous only at and above some maximum feature size fixed at the end of the transition.  Due to statistical variation, the process also need not appear isotropic from any particular vantage point, particularly if the number of observable nucleation events is small, as would tend to be the case early in the transition.  As far as the gravitational field is concerned, however, any induced inhomogeneities should be far less severe than those already existing due to the presence of voids and superclusters in the universe, which involve essentially the entire dust contribution to stress-energy, while we will model effects involving $1\%$-$5\%$ of the cosmic dust.  Moreover, the effect of mass consumption will be to take mass from the existing overdense regions of the universe and convert it into waste radiation, which should relatively quickly become evenly distributed throughout the universe.

\section{Construction of the Framework}

For reasons discussed in section 2, we will restrict our analysis to constant-velocity expansion, i.e.  $V_{i}(t',t) =\frac{4 \pi}{3} (\int_{t'}^{t} \frac{v_{i} \, \theta(t''-t')}{a(t'')} dt'')^3 $, though we note that other choices (such as constant coordinate-velocity in a low-velocity regime) could equally well be analyzed, corresponding to different and perhaps less aggressive behavior assumptions.  Similarly, we will always use an appearance rate $f(t)$ of the form described in section 2.  The accessibility of cosmic dust is assumed to be a constant for all behavior types.  The resource consumption rate amongst the various strategies is specified by the decay constants $\Gamma_{i}$ and the saturation time by $T_i$.  Thus, for our purposes, a complete expansion and resource consumption strategy is given by the constants $s_i = \{v_i,T_i,\Gamma_i \}$, while a complete scenario is given by specifying all the $s_i$ together with the appearance and mass accessibility constants $\alpha_i$ and $A$.  We now describe how the framework is constructed.

\subsection{Multiple expansion strategies }

Instead of working with the unsaturated fraction of space, $g(t)$, as in the last section, we will often find it convenient to work with the fraction of space saturated by life of behavior type $i$, namely $h_{i}(t)$, and the total saturated fraction $h(t) = \sum_{i} h_{i}(t) = 1 - g(t)$.  We thus express the real, average saturated volume function in terms of $h(t)$ as:
\begin{widetext}
\begin{eqnarray}
\bar{V_i}(t',t) = \frac{1}{1-h(t')} \int_{t'}^{t} dt''(1-h(t'')) \frac{d V_{i}(t',t''-T_i)}{dt''} \\
 = \frac{4 \pi  v_i  }{1-h(t')} \int_{t'}^{t} dt'' (1-h(t''))\theta(t'' - t'-T_i) \frac{ \left(\int_{t'}^{t''-T_i} \frac{v_i \theta(t''' - t')}{a(t''')} \, dt'''\right){}^2}{a(t''-T_i)} \\
 = \frac{6^{2/3} \ \pi^{1/3} \  v_i }{1-h(t')}  \int_{t'}^{t} dt'' (1-h(t'')) \frac{V_{i}(t',t'' -T_i)^{2/3} }{a(t''-T_i)}.
\end{eqnarray}
\end{widetext}
As noted in the previous section, for a single strategy with $T=0$, when this volume function is substituted into the CK integral equation, one obtains the GTW formula as a solution.  Here we are interested in the case of $T_i \neq 0$ and multiple strategies, for which we utilize a coupled set of analogous integral equations, namely:
\begin{eqnarray}
h_{i}(t) = \int_{0}^{t} dt' \ (1 - h(t')) \ f_{i}(t') \ \bar{V_i}(t',t).
\end{eqnarray} 
One should be careful to note the sum $(1 - h(t')) = (1 -\sum_{j} h_j(t'))$ that is lurking under the integral.  For numerical calculations, it is convenient to transform these into a coupled set of first-order delay differential equations, giving:
\begin{eqnarray}
\dot{h_i}(t) &=&  (1-h(t)) \frac{6^{2/3} \pi^{1/3} v_i }{a(t-T_i)} X_i(t-T_i) \\
\dot{X_i}(t) &=& \frac{2^{5/3} (\frac{\pi}{3})^{1/3} v_i }{a(t)} Y_i(t) \\
\dot{Y_i}(t) &=& \frac{2^{2/3} (\frac{\pi}{3})^{1/3} v_i }{a(t)} Z_i(t) \\
\dot{Z_i}(t) &=& f_i(t).
\end{eqnarray}
Here an overdot has denoted a time derivative, and the functions $X(t)$, $Y(t)$, and $Z(t)$ are identified by:
\begin{eqnarray}
X_i(t) &=& \int_{0}^{t} dt' \ f_i(t') \ V_i(t',t)^{2/3} \\
Y_i(t) &=& \int_{0}^{t} dt' \ f_i(t') \ V_i(t',t)^{1/3} \\
Z_i(t) &=& \int_{0}^{t} dt' \ f_i(t').
\end{eqnarray} 

In the limit that life has zero noticeable backreaction on the universe, these will be sufficient to obtain the saturated fractions $h_i(t)$ -- in the case that life transforms mass into radiation, we will need to couple these to the Friedmann equations to simultaneously solve for the scale factor $a(t)$.

\subsection{Coupling to the Friedmann equations}

First, we divide the pressureless dust $\rho_m$ into two components, an ``accessible" component and an ``inaccessible" one, i.e. $\rho_{m_A} = A \rho_m$ and $\rho_{m_I} = I \rho_m$ with $A$ and $I$ constants satisfying $A+I=1$.  Each component obeys the standard equations of motion for dust, namely $\frac{d}{dt} (a^3 \rho_{m_A})=0$ and $\frac{d}{dt} (a^3 \rho_{m_I})=0$.  With the onset of aggressively expanding life of $n$ distinct strategies, we further subdivide $\rho_{m_A}$ into $n+1$ components $\rho_{m_i}$, with $i$ ranging from $0$ to $n$.  We take the $0$-component to represent the unsaturated fraction of $\rho_{m_A}$, i.e. $\rho_{m_0}=g \rho_{m_A}$ while the $i^{\textrm{th}}$ component is that saturated by the $i^{\textrm{th}}$ strategy type, i.e. $\rho_{m_i}=h_i \rho_{m_A}$.  On a sufficiently large scale, then, we write the $00$-component of the stress-energy tensor of the universe as $\rho = \rho_\Lambda + \rho_r + \rho_{m_I} + \sum_{i=0}^{n} \rho_{m_i}$, where the $\Lambda$  and $r$ subscripts denote the energy density of the cosmological constant and radiation, respectively.

In the absence of any mass consumption, the time-dependence of the $h_i$ cause the $\rho_{m_i}$ to evolve according to:
\begin{eqnarray}
\frac{d}{dt} ( \rho_{m_0} a^3) = \dot{g} \ (\rho_{m_a} a^3) = \frac{\dot{g}}{g} \ (\rho_{m_0} a^3)  \\
\frac{d}{dt} (\rho_{m_i} a^3) = \dot{h}_i \ (\rho_{m_a} a^3) = \frac{\dot{h}_i}{g} \ (\rho_{m_0} a^3).
\end{eqnarray}
If we now add the assumption that life converts mass into radiation at a rate proportional to the total amount of accessible mass it controls, the above equations are modified to:
\begin{eqnarray}
\frac{d}{dt} ( \rho_{m_0} a^3) &=& \frac{\dot{g}}{g} \ (\rho_{m_0} a^3)  \\
\frac{d}{dt} ( \rho_{m_i} a^3) &=& \frac{\dot{h}_i}{g} \ (\rho_{m_0} a^3) - \Gamma_i (\rho_{m_i} a^3). 
\end{eqnarray}
The FRW continuity equation $\dot{\rho} = -3 \frac{\dot{a}}{a}(\rho + p)$ then implies, on a sufficiently large scale, that the radiation component evolves according to:
\begin{eqnarray}
\frac{d}{dt} ( \rho_{r} a^4) &=& \sum_{i=1}^{n} \Gamma_i (\rho_{m_i} a^4).
\end{eqnarray}

These equations, combined with the first Friedmann equation $(\frac{\dot{a}}{a})^2 = \frac{8 \pi G \rho + \Lambda}{3}$ and those of the previous section give a complete set for modeling the saturation of the universe by aggressive civilizations, incorporating the physical backreaction on the universe itself.  We are now in a position go further and address questions of how the backreaction, observability, relative success of different strategies, and first-mover advantage will be quantified and plotted based on this model.

\subsection{Backreaction}

Having coupled the Friedmann equations to the waste heat generated by advanced life, we see that a cosmological backreaction appears in this model, which will depend strongly on resource consumption. In the conversion of mass to radiation, energy density is conserved (though the energy density of radiation will more quickly be diminished with the expansion of the universe), but pressure will be abruptly increased.  Due to the Friedmann acceleration equation $\frac{\ddot{a}}{a} = -\frac{4 \pi G}{3}(\rho + 3p) + \frac{\Lambda}{3}$ which is implied by the first Friedmann equation and the continuity equation in the previous subsection, one can see that the effect of converting mass to radiation will be to temporarily slow the expansion of the universe until the cosmological constant eventually dilutes all gravitational effects of matter and causes the universe to asymptotically approach a de Sitter solution.  We will quantify this effect by plotting the fractional change in the scale factor and the Hubble parameter, i.e. $\frac{\Delta a(t)}{a(t)} = \frac{a_{\mathrm{life}}(t)-a_{\mathrm{no\ life}}(t)}{a_{\mathrm{no\ life}}(t)}$ and $\frac{\Delta H(t)}{H(t)}= \frac{H_{\mathrm{life}}(t)-H_{\mathrm{no\ life}}(t)}{H_{\mathrm{no\ life}}(t)}$, where the subscript ``no life" denotes the standard cosmology solution and ``life" is the solution from our model.  In the next section, we will survey a range of models leading to maximum fractional changes of order $10^{-5}$ to $10^{-3}$. 

\subsection{Average domain size, first-mover advantage, and relative strategy success}

Perhaps the most obvious way to compare the relative success of different strategies and starting-times is to compute the ratio of their final average coordinate volumes, $\lim\limits_{t' \rightarrow \infty} \bar{V_i}(t,t')$.  That is, the first mover advantage of a species of strategy $i$ that appears at time $t$, relative to another of the same strategy that appears at a later time $t'$ can be expressed by
\begin{eqnarray}
A_i(t,t') \equiv \lim\limits_{t'' \rightarrow \infty} \frac{\bar{V_i}(t,t'')}{\bar{V_i}(t',t'')}.
\end{eqnarray}
Similarly, comparing the success of strategy $i$ to strategy $j$ for equal appearance time $t$ can be expressed by
\begin{eqnarray}
A_{i j}(t) \equiv \lim\limits_{t' \rightarrow \infty} \frac{\bar{V_i}(t,t')}{\bar{V_j}(t,t')}.
\end{eqnarray}
One may of course also compare differing strategies at different starting times with a ratio $A_{i j}(t,t')$ in a completely analogous way.

We should note in passing that the term ``success" in this context is rather loaded -- we do not mean to suggest a universal measure of utility, but only a simple means of comparing average domain size across different starting times and strategies.

\subsection{Prior and conditional probabilities of observability}

For a given scenario, two quantities of interest are the prior probability $P(\mathrm{observable})$ that a randomly chosen point in space (at a given time) is within the future light-cone of saturation taking place somewhere in the universe, and the conditional probability $P(\mathrm{observable}|\mathrm{not\ saturated})$ that a point in space is in the future light-cone of a saturation event, \emph{given that the point has not itself been saturated}.  To see how these are related, we note the following:
\begin{widetext}
\begin{eqnarray}
P(\mathrm{observable}) &=& P(\mathrm{observable}|\mathrm{not\ saturated})P(\mathrm{not\ saturated}) \nonumber \\
 &+& P(\mathrm{observable}|\mathrm{saturated})P(\mathrm{saturated}).
\end{eqnarray}
Furthermore, we assume that $P(\mathrm{observable}|\mathrm{saturated}) = 1$, i.e. we are discounting the possibility of a locally invisible saturation.  This, combined with the normalization condition $P(\mathrm{not\ saturated}) + P(\mathrm{saturated}) = 1$ allows us to conclude the following:
\begin{eqnarray}
P(\mathrm{observable}|\mathrm{not\ saturated}) = 1 - \frac{P(\mathrm{not\ observable})}{P(\mathrm{not\ saturated})}.
\end{eqnarray}

\end{widetext}
Now, we already know $P(\mathrm{not\ saturated})$ -- it is just $g(t)$.  Thus, we only need to compute $P(\mathrm{not\ observable})$.

To calculate $P(\mathrm{not\ observable})$, which is the fraction of space \emph{not} within the future light cone of a saturation event, which we denote as $j(t)$, two cases are worth treating separately.  We treat here the case in which all saturation fronts advance with subluminal speed in the co-moving frame, and discuss the case of superluminal propagation in the next section.

The light cones of various strategies are not mutually exclusive, unlike saturated domains.  In fact, since light cones of all strategies expand at the same speed of light, we can use the technique of virtual bubbles and independent events, so that $j(t) = \prod_{i} j_{i}(t)$.  This implies that a GTW-like formula holds for each strategy, i.e.
\begin{eqnarray}
j_i(t) = e^{- \int_{0}^{t-T_i}dt' f_i(t')  V^{v=1}(t'+ T_i,t) }
\end{eqnarray}
where $V^{v=1}$ is the standard volume function we have been using for the specific case of $v=1$ (light), and the upper limit of integration is due to the fact that $V(t',t)$ vanishes for $t'>t$.

\begin{widetext}
\subsection{Superluminal propagation of the saturation front}

In the case that the probe front is sufficiently fast, and the saturation time $T$  is sufficiently long, it is possible that the saturation front advances at a velocity faster than light.  If the velocity of the probe front in the co-moving frame is a constant $v$ (the parameter in our volume functions), then the velocity of the saturation front in the co-moving frame is given by a time-dependent $v_s$ according to:
\begin{eqnarray}
v_s(t) = v \frac{a(t)}{a(t-T)}.
\end{eqnarray} 
In the standard cosmology, this is a decreasing function which asymptotically approaches a constant as the universe approaches a de Sitter solution in the distant future.  If all saturation fronts remain superluminal, then the situation is straightforward, $j(t)=g(t)$ and $P(\mathrm{observable}|\mathrm{not\ saturated})=0$ -- i.e. the first observable sign that the universe is being saturated by life, in such a scenario, is to see one's own neighborhood become saturated.  In the case of a single-strategy scenario which undergoes a transition from superluminal to subluminal saturation fronts, we can again use the technique of virtual bubbles and independent events, combined with the appropriate volume function given by:
\begin{eqnarray}
V^{obs}(t',t)   =   \frac{4 \pi}{3} \left( \int_{t'+T}^{t_{trans}}dt'' \frac{v}{a(t''-T)} + \int_{t_{trans}}^{t}dt'' \frac{1}{a(t'')}\right)^3 \\
\hspace*{-2.2cm}  =   \frac{4 \pi}{3}  \left( \int_{t'}^{t-T}dt'' \left[ \theta(\frac{v \, a(t''+T)}{a(t'')} - 1) \frac{v}{a(t'')} + \theta(1 - \frac{v \, a(t''+T)}{a(t'')}) \frac{1}{a(t''+T)}  \right] \right)^3.
\end{eqnarray}
That is, while the saturation fronts advance superluminally, the observability volume is just the saturation volume itself.  When the saturation fronts make the transition to subluminal velocity at $t_{trans}$ (the solution to $ v \frac{a(t_{trans})}{a(t_{trans}-T)} = 1$), the observability volume continues to grow at the speed of light.  The single-strategy assumption then gives rise to a formula for $j(t)$ in GTW form:
\begin{eqnarray}
j(t) = e^{- \int_{0}^{t-T}dt' f(t')  V^{obs}(t',t) }.
\end{eqnarray}
This is in agreement with the result for the purely subluminal case in the previous section (for the case of a single strategy), due to the definition of $V^{obs}$.

In the general case of many strategies with saturation fronts making the superluminal/subluminal transition at various times, the geometry is less clear -- we cannot invoke virtual bubbles and independent events (due to the multiple expansion velocities of the observability volumes), nor can we invoke mutual exclusivity, due to the overlapping of light-cones.  Thus, we leave the most general case for $j(t)$ unsolved at this time.  The strategy parameters required for superluminal propagation of the saturation front are rather extreme -- we plot in figure 3 the parameters $v$ vs. $T$ required to satisfy $v_s = 1$ at various universe ages.  Even at the time of the earliest possible civilizations (7.5 Gyr), and assuming a long saturation time of half a billion years, one can see that $v_s =1$ requires a probe front velocity of $v \approx .95 c$ in the co-moving frame (for a Lorentz gamma factor of $\gamma \approx 3.2$).  One can see that the numbers become more extreme at later cosmic times.

\begin{figure}
\centering
\includegraphics[width=0.55\linewidth]{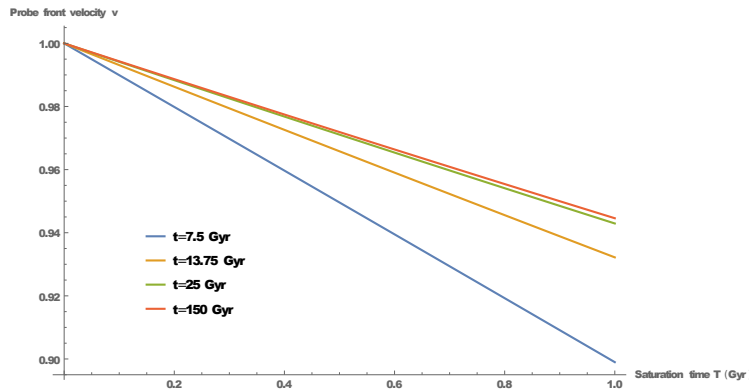}
\caption{Probe front velocity $v$ vs. saturation time $T$ required to achieve a saturation front velocity of $v_s = 1$ at various cosmic times $t$.  The area above any line represents parameter pairs resulting in superluminal propagation of the saturation front.  This is easier to achieve at early cosmic times, and eventually reaches a final maximum difficulty at late times as $a(t)$ approaches a de Sitter solution.}
\label{Figure1}
\end{figure}

\section{A Survey of example scenarios}
We now explore a range of scenarios to graphically illustrate some basic features of this process.  As a convenient way to do this, we will first describe a ``default scenario" as a baseline from which to vary our parameters (it is not to be regarded as a ``most probable scenario," as so little is known about the input parameters -- it simply represents one scenario with convenient timescales).  All graphs involving functions of the energy density are expressed in units of the current critical density of the universe as calculated from the standard baseline cosmology we employ, $\rho_{\mathrm{crit}}(13.75)$.  We take the default scenario to be a single-strategy scenario with $ \left\lbrace  v,T,\Gamma,A,\alpha \right\rbrace = \left\lbrace .5,.1 \ Gyr,\frac{\ln(2)}{10 \ Gyr},.025,\frac{1}{100 \ Gly^3 Gyr} \right\rbrace$ -- that is, all aggressively expanding life expands with a probe front moving at .5c, takes 100 million years to saturate matter after the initial wave of probes have passed by, converts saturated matter to radiation with a half-life of ten billion years, is able to saturate $2.5\%$ of the cosmic dust within the volume it controls, and such life is currently appearing in the universe at a rate of $\approx 1$ appearance per 100 billion cubic light years of unsaturated coordinate volume, per billion years of cosmic time.  The scenario is depicted in figure 4.  In this scenario, the universe is presently $\approx 88 \%$ saturated, and saturation became visible ``almost everywhere" ($P(\textrm{obs}) = .999$) at $t \approx 12.5 \, Gyr$. Another figure perhaps of interest is the average final coordinate volume occupied by such a species appearing at present day ($t = 13.75 \, Gyr$), namely $0.31 \, Gly^3$ -- for comparison purposes, the volume of the Virgo supercluster is roughly $10^{-3} \, Gly^3$, and the final coordinate volume encompassed by a spherical light pulse emitted today is $ 2 \times 10^4 \, Gly^3$.

Also notable is the magnitude of first-mover advantage (a very early civilization can expect to capture multiple hundreds of times more volume than one appearing at present day) and the backreaction on the evolution of $a(t)$ and $H(t)$, resulting in maximum changes on the order of $10^{-4}$ -- note that $\frac{\Delta H}{H}$ eventually returns to 0, as the future evolution of the universe is increasingly dominated by the cosmological constant, while the scale factor remains permanently smaller than it would otherwise be.  Using this scenario as a standard, we now vary parameters in order to point out some specific features of interest.

\begin{figure}
\subfloat[]{
  \includegraphics[width=0.45\linewidth]{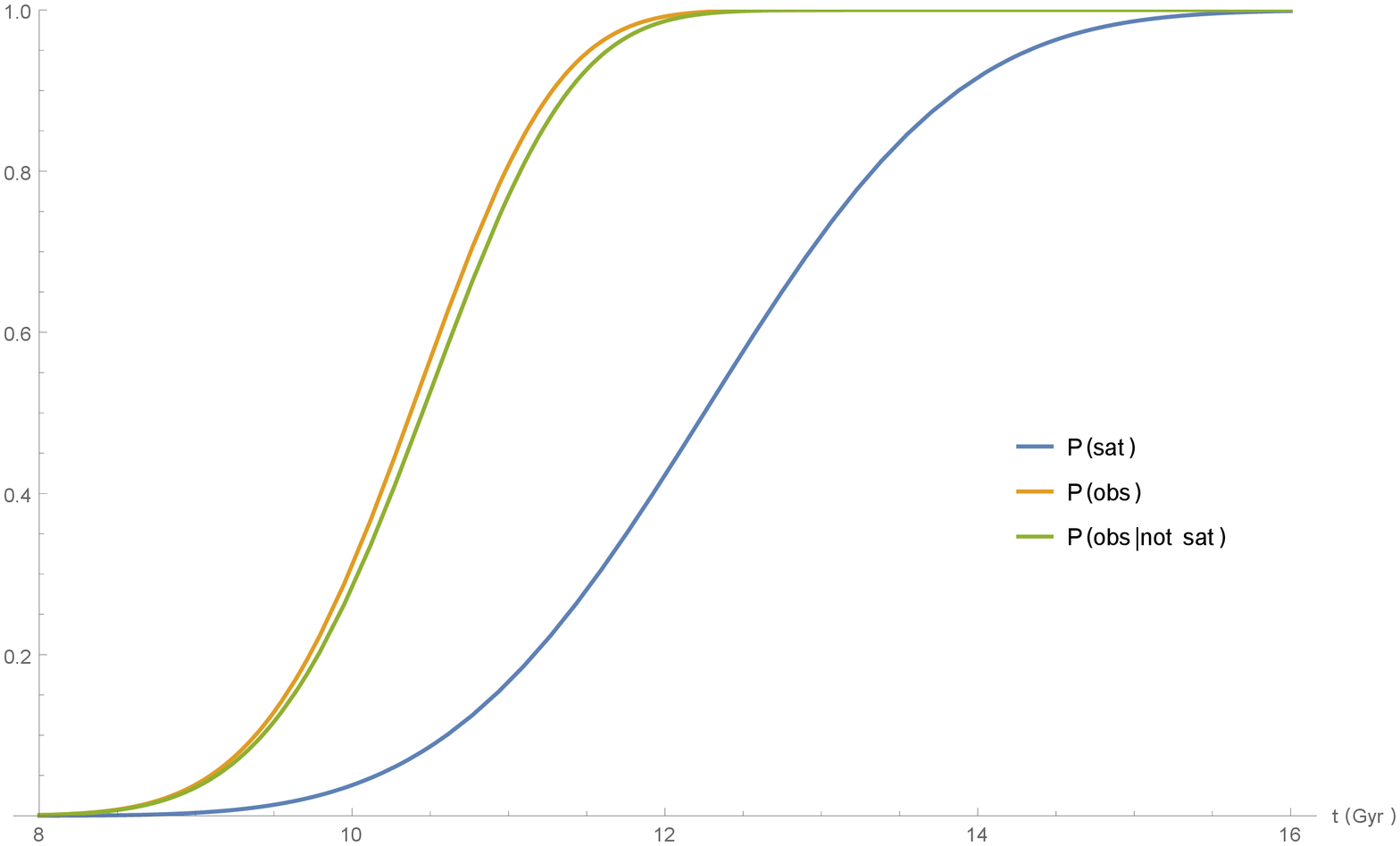}
}
\subfloat[]{
  \includegraphics[width=0.45\linewidth]{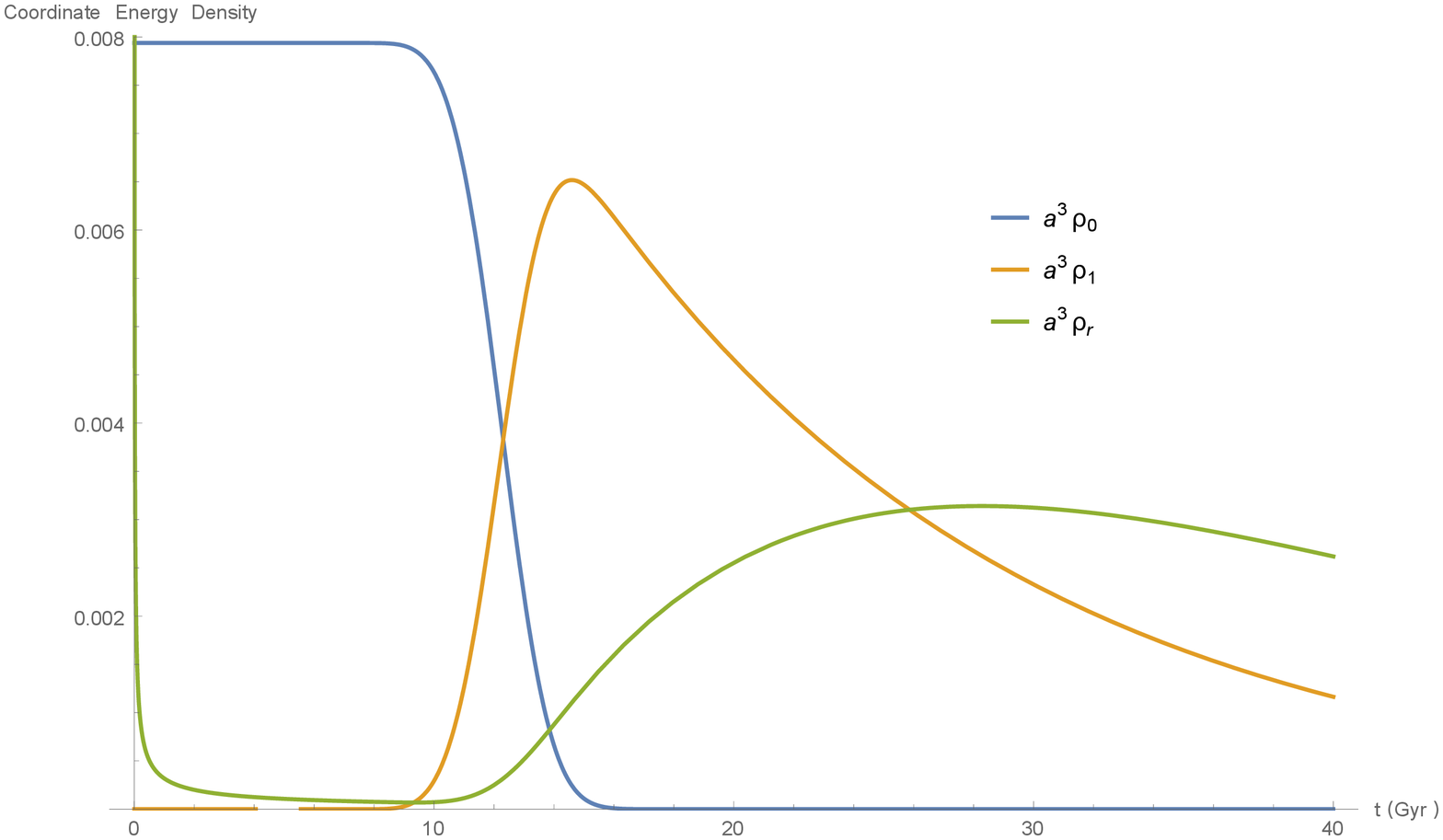}
}
\hspace{0mm}
\subfloat[]{
  \includegraphics[width=0.45\linewidth]{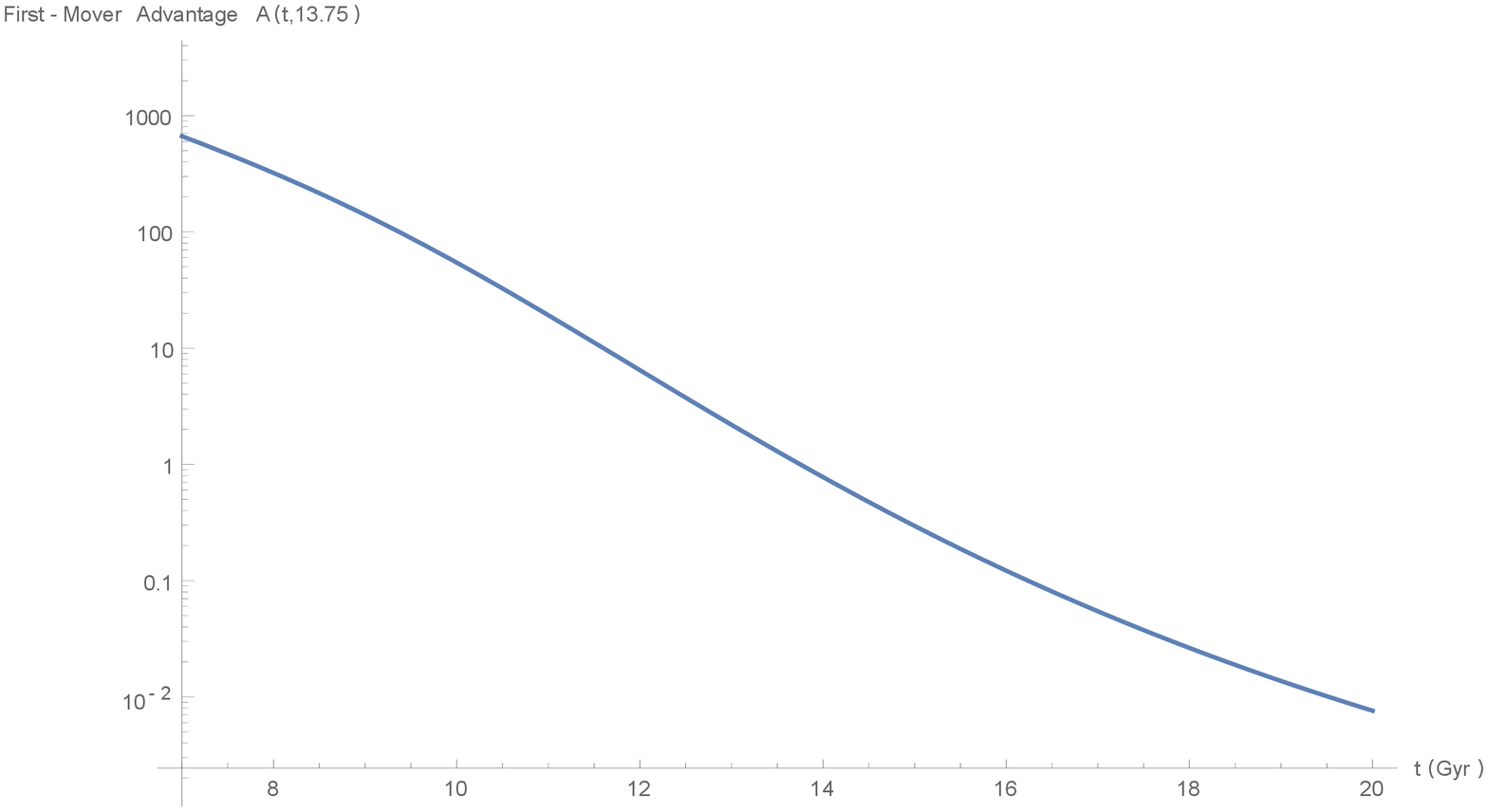}
}
\subfloat[]{
  \includegraphics[width=0.45\linewidth]{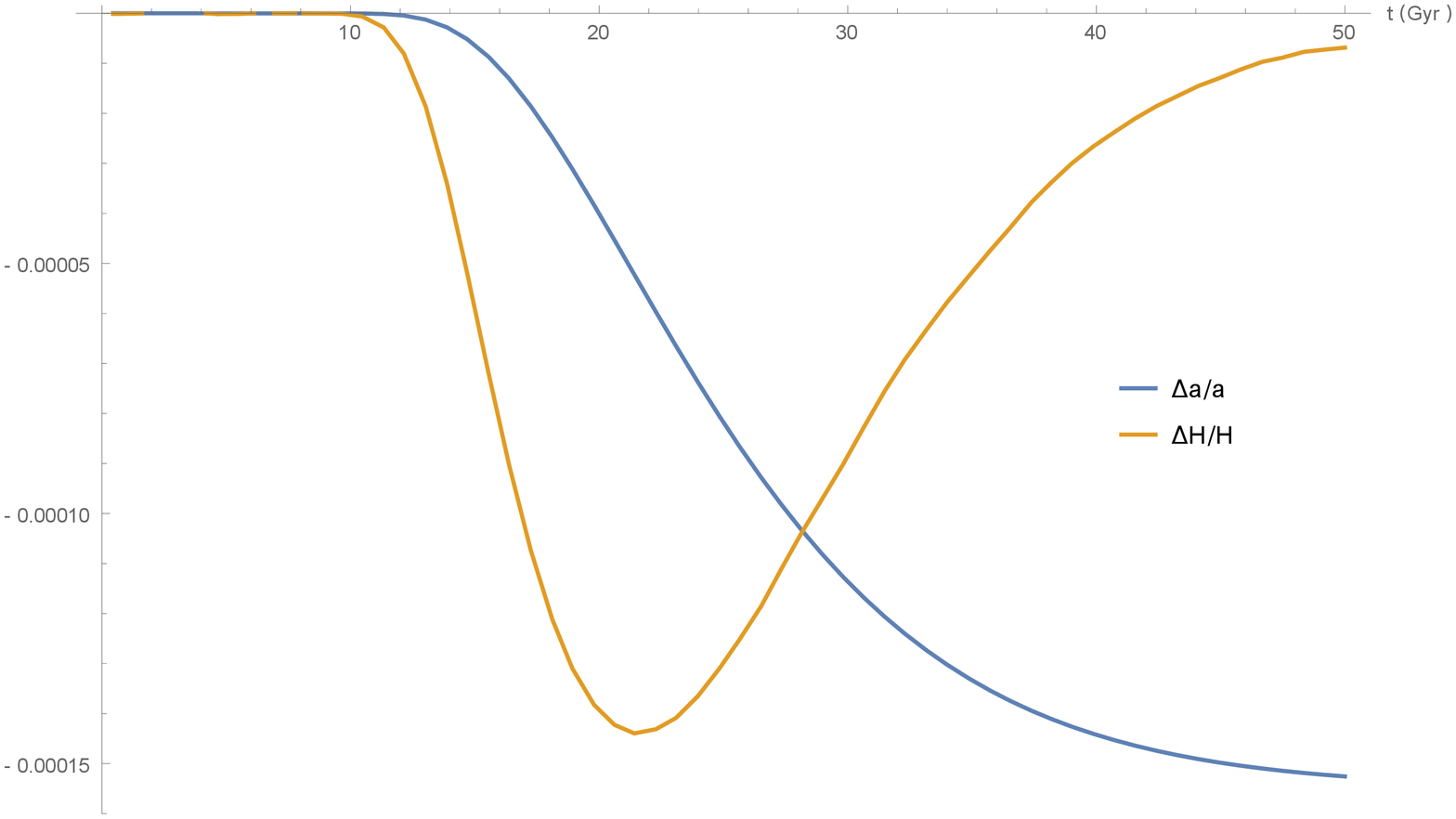}
}
\caption{The default scenario with $ \left\lbrace  v,T,\Gamma,A,\alpha \right\rbrace = \left\lbrace .5 \, c,.1 \ Gyr,\frac{\ln(2)}{10 \ Gyr},.025,\frac{1}{100 \ Gly^3 Gyr} \right\rbrace$. (a) illustrates the probability that a random point in space is saturated $P(\textrm{sat})$, the fraction of space from which saturation is observable $P(\textrm{obs})$, and the conditional probability that a non-saturated point can observe saturation, $P(\textrm{obs} | \textrm{not sat})$.  (b) illustrates the ``coordinate energy density" ($a^3 \rho$) of unsaturated but accessible matter $\rho_0$, saturated matter $\rho_1$, and radiation $\rho_r$ -- energy density is expressed in units of the current critical density of the universe, $\rho_{\textrm{crit}}(13.75)$.  (c) illustrates the first-mover advantage function $A(t,13.75)$ relative to an aggressively expanding species appearing today.  (d) illustrates the fractional change (slowing) in the scale factor and Hubble parameter induced by the conversion of mass to waste radiation.   }
\end{figure}

\subsection{Superluminal propagation of the saturation front  }
We describe the single-strategy scenario $ \left\lbrace  v,T,\Gamma,A,\alpha \right\rbrace = \left\lbrace .94,.8 \ Gyr,\frac{\ln(2)}{10 \ Gyr},.025,\frac{1}{200 \ Gly^3 Gyr} \right\rbrace$ in figure 5.  Compared to the default scenario, we have made the probe front much faster, the saturation time much longer, and cut the appearance rate by a factor of 2 (mass consumption parameters are left unchanged).  The effect is to induce a superluminal/subluminal transition in the saturation front that takes place at approximately $11.5 \,Gyr$ -- $P(\mathrm{obs}|\mathrm{not \ sat})$ is exactly zero before this transition and climbs slowly after it.  Equivalently, one can see that $P(\mathrm{sat})$ and $P(\mathrm{obs})$ are essentially on top of each other throughout the duration of the transition.

In this scenario, the universe is currently 98.8\% saturated, but saturated matter is only visible from 2.5\% of the small remaining unsaturated portion of the universe.  Such models illustrate the complexity in interpreting a potential null result in searches for intergalactic saturated matter.

\begin{figure}
\centering
\subfloat[]{
  \includegraphics[width=0.45\linewidth]{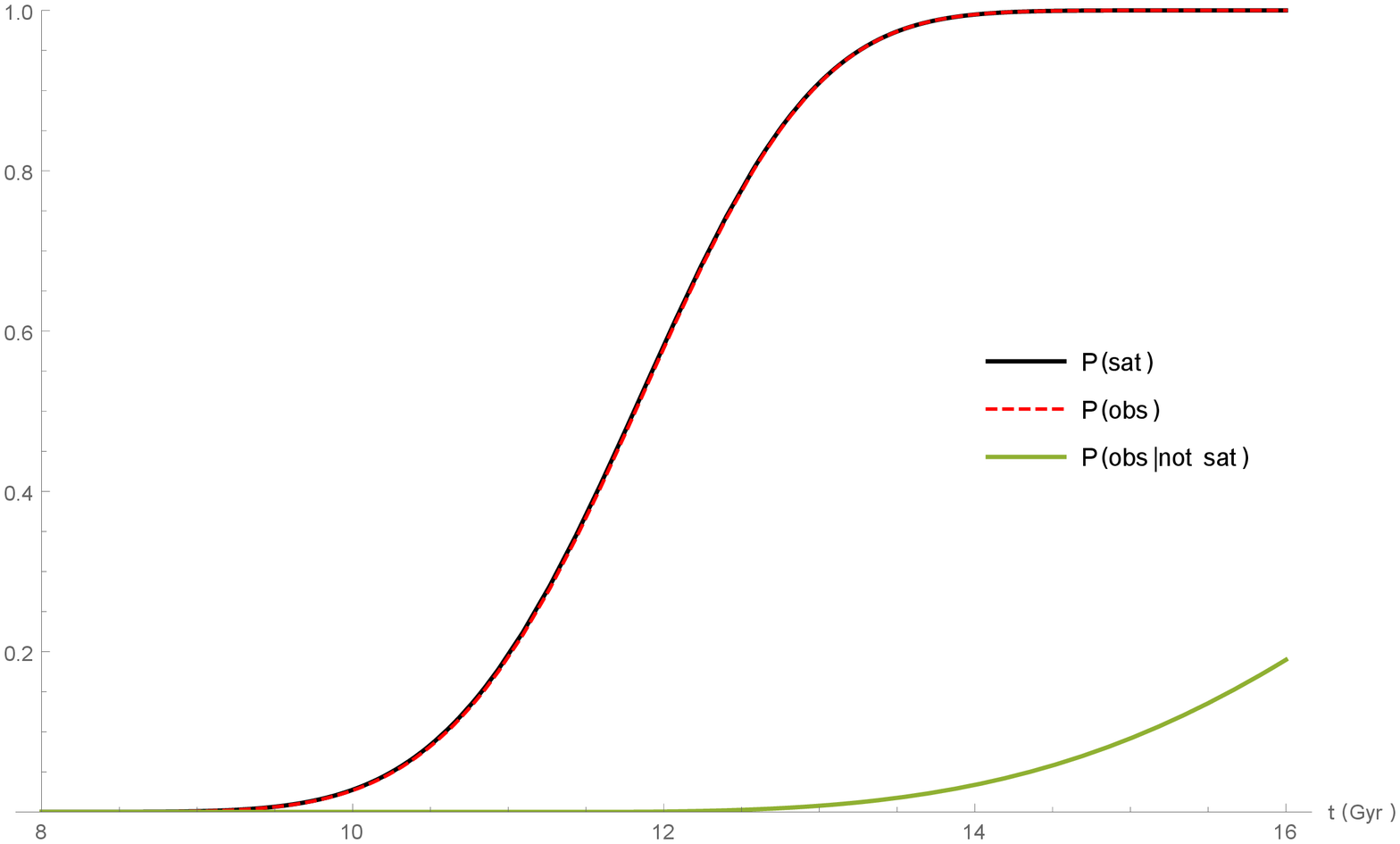}
}
\subfloat[]{
  \includegraphics[width=0.45\linewidth]{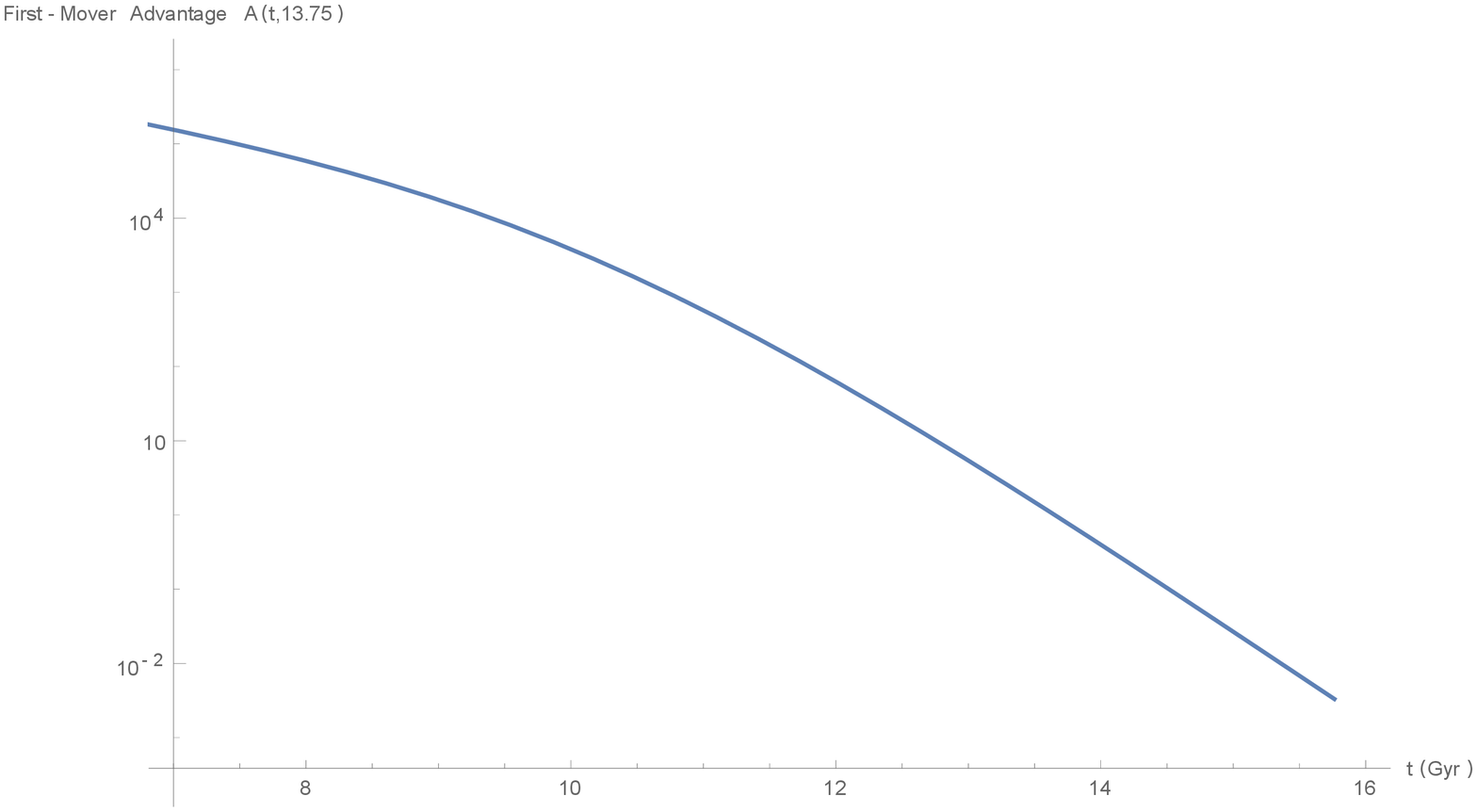}
}
\caption{Superluminal scenario with $ \left\lbrace  v,T,\Gamma,A,\alpha \right\rbrace = \left\lbrace .94,.8 \ Gyr,\frac{\ln(2)}{10 \ Gyr},.025,\frac{1}{200 \ Gly^3 Gyr} \right\rbrace$. Note that in (a), the saturation takes place much more quickly than the default scenario, but $P(\textrm{obs} | \textrm{not sat})$ increases much more slowly. }
\end{figure}

\subsection{Rare fast expanders vs. common slow expanders  }

Here we introduce our first competitive scenario with two distinct expansion strategies.  The scenario is defined by $ \left\lbrace  v_1,v_2,T_1,T_2,\Gamma_1,\Gamma_2,A,\alpha_1,\alpha_2 \right\rbrace = \left\lbrace .5c,.01c,.1 \, Gyr,0 \, Gyr,\frac{\ln(2)}{10 \ Gyr},\frac{\ln(2)}{10 \ Gyr},.025,\frac{1}{100 \, Gly^3 Gyr},\frac{100,000}{100 \, Gly^3 Gyr} \right\rbrace$ and is depicted in figure 6.  We have taken the default scenario and added a second type of civilization that is far more common, but expands far more slowly. The zero saturation time of the second behavior type could be interpreted as characteristic of a strategy in which successive outward steps are taken only after the local matter has been fully saturated -- this would also be consistent with the much slower probe front.  

Note that, because of the models used for the appearance rate and expansion volume, the equations for saturated fraction $h_i(t)$ depend on the multiplicative combination $\alpha_i \, v_{i}^3$, so that the dynamics of $h_i(t)$ is invariant under scalings that leave this combination constant.  For example, $h_{i}(t)$ will be unchanged if $v_i$ is decreased by a factor of 10 and $\alpha_i$ is simultaneously increased by a factor of 1,000.  Using this fact, it is easy to generate a variety of competitive strategy scenarios covering many orders of magnitude in $v$ and $\alpha$.  Note, however, that in the general case neither $P(\mathrm{obs})$ nor $P(\mathrm{obs} | \mathrm{not \, sat})$ is invariant under such a scaling -- as figure 6b illustrates, the presence of a ``slow but common" behavior type in the universe causes saturated matter to very rapidly become observable from almost everywhere. 

\begin{figure}
\centering
\subfloat[]{
  \includegraphics[width=0.45\linewidth]{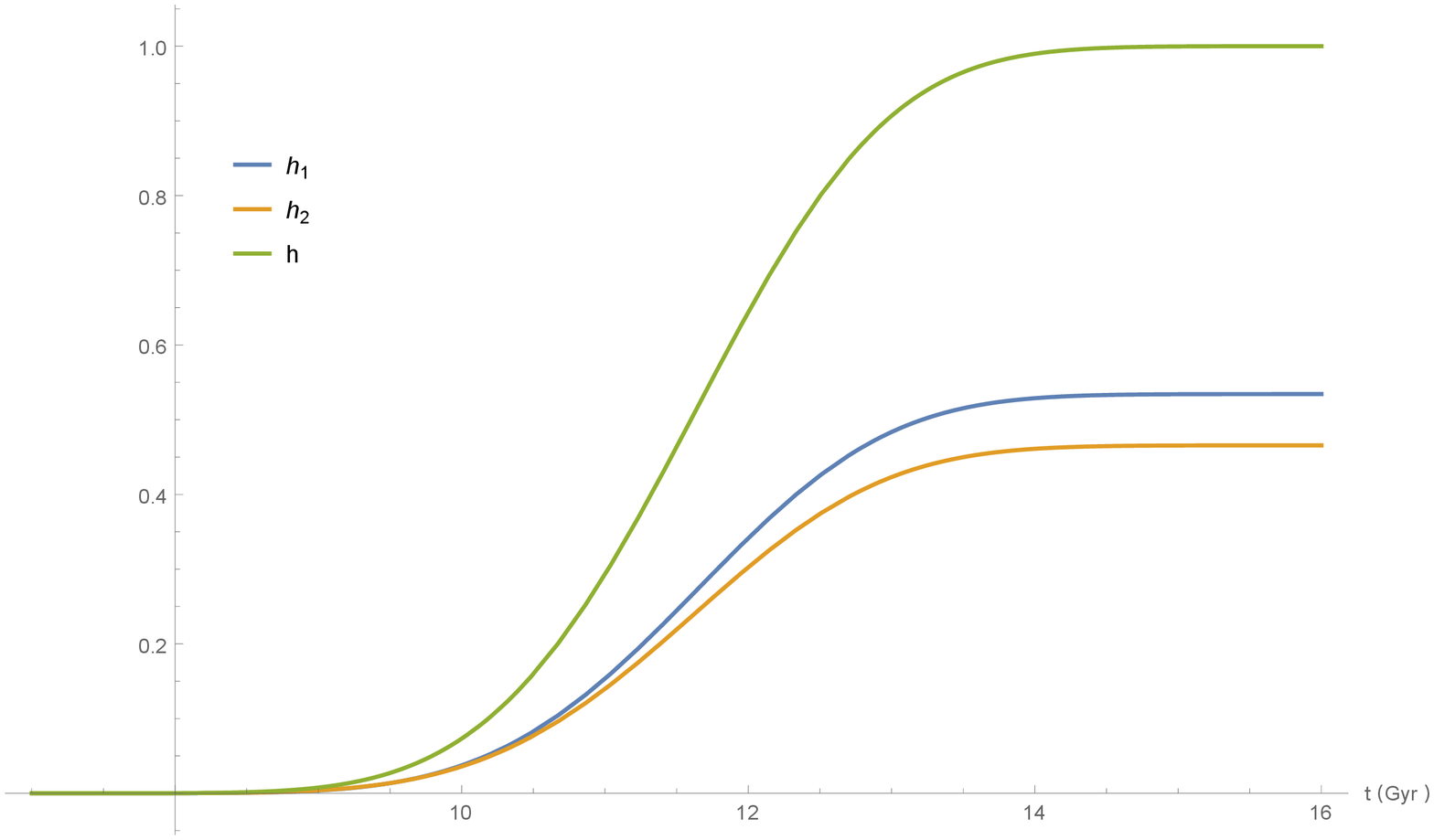}
}
\subfloat[]{
  \includegraphics[width=0.45\linewidth]{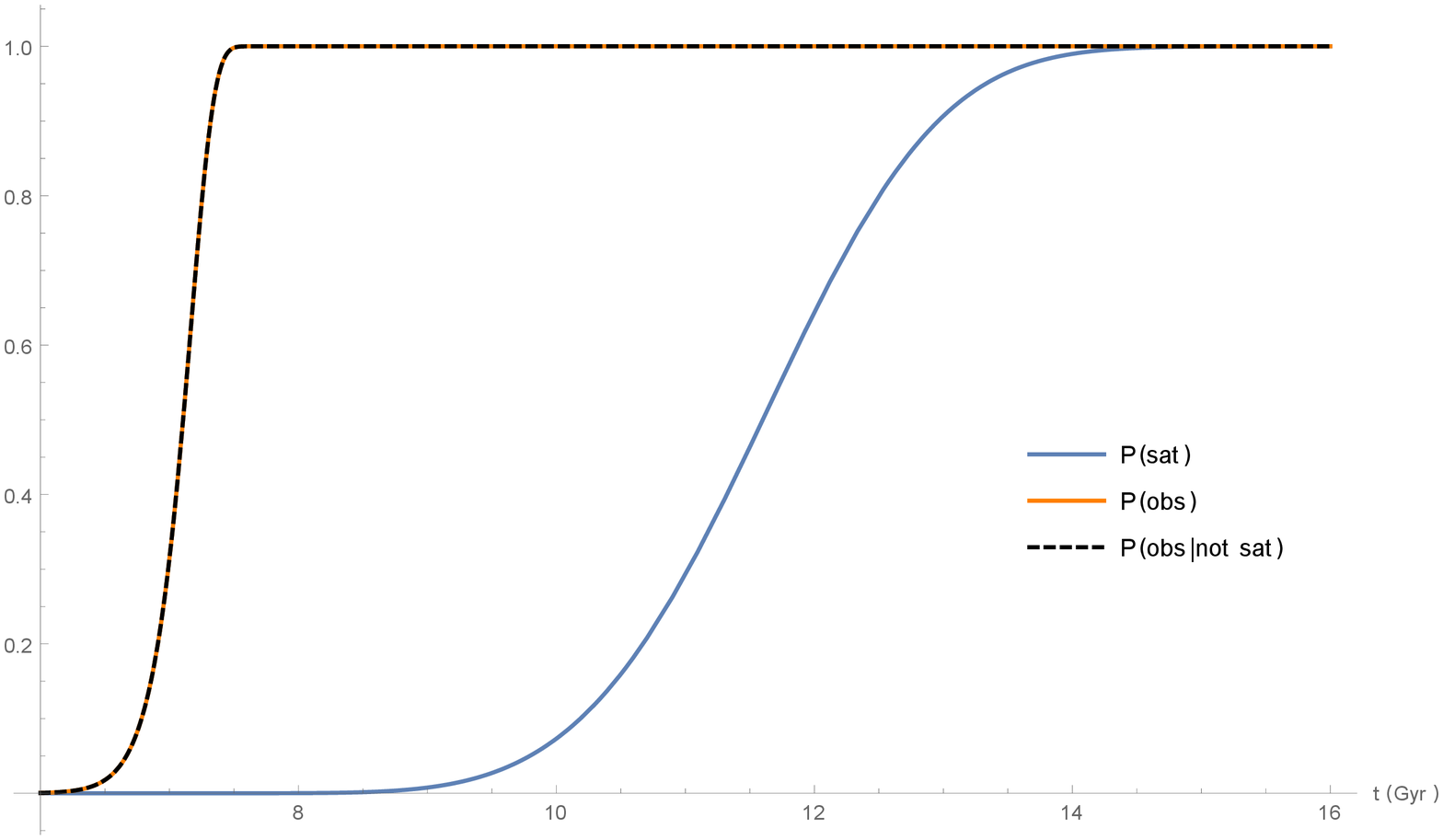}
}
\hspace{0mm}
\subfloat[]{
  \includegraphics[width=0.45\linewidth]{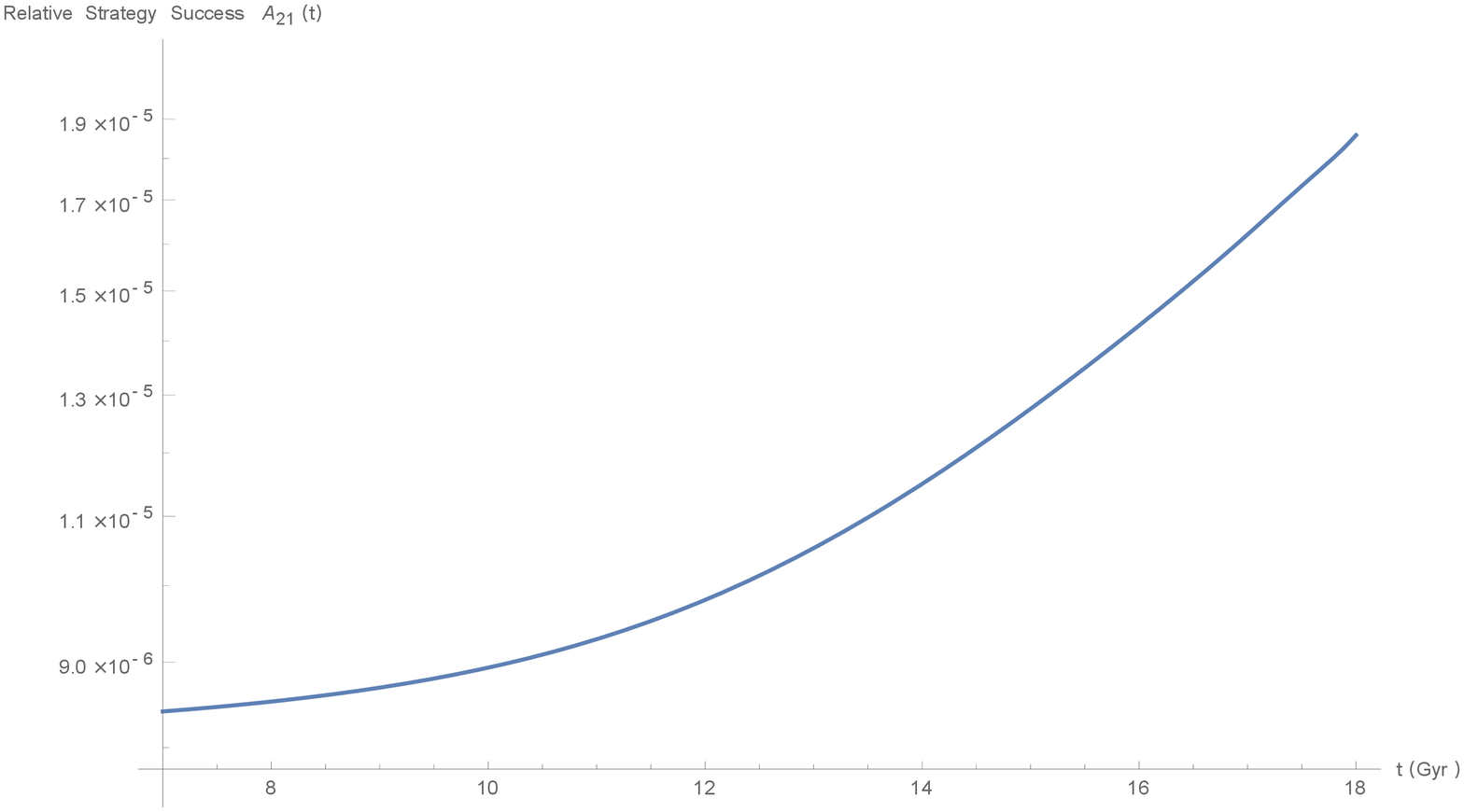}
}
\subfloat[]{
  \includegraphics[width=0.45\linewidth]{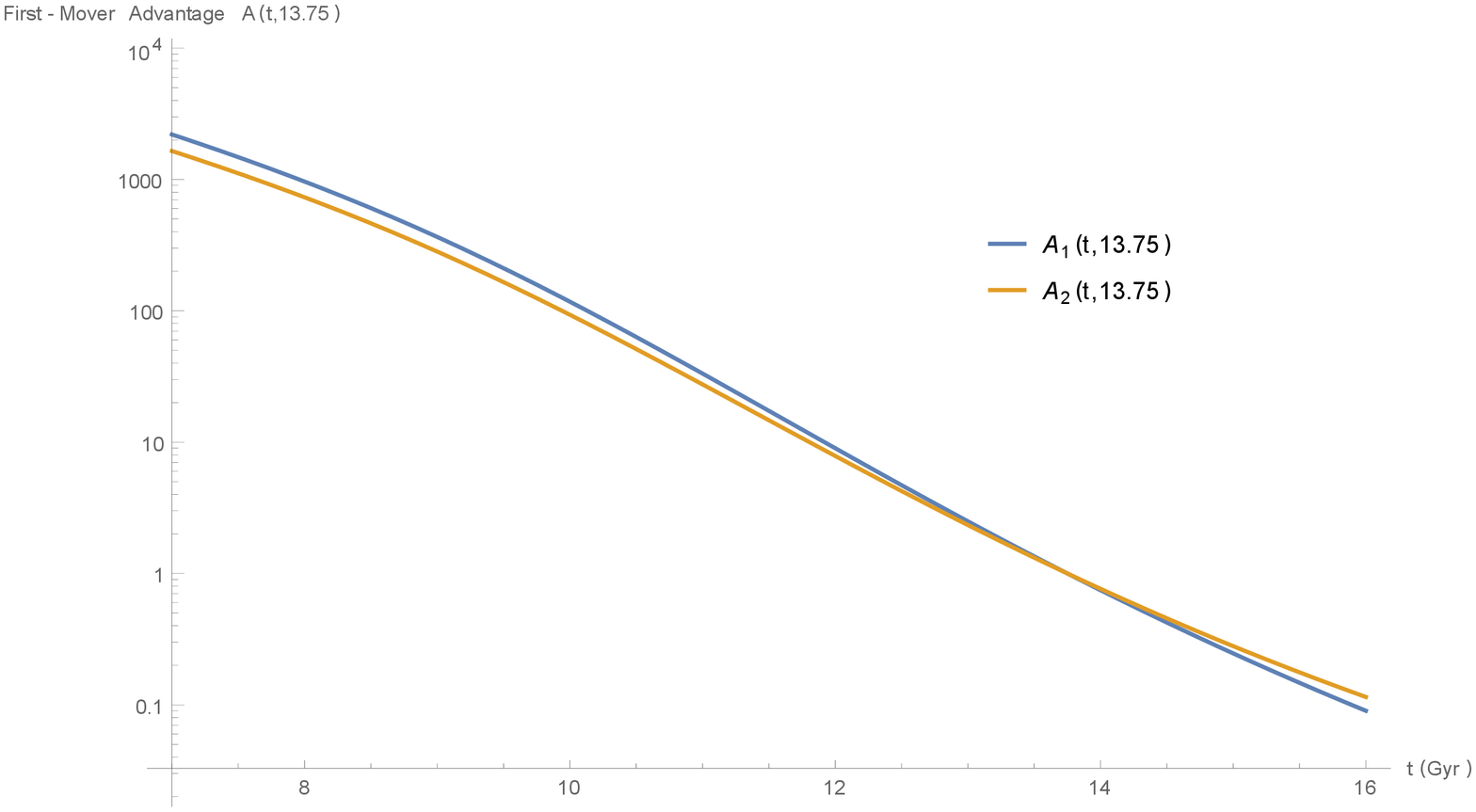}
}
\caption{Fast vs. common scenario described by $ \left\lbrace  v_1,v_2,T_1,T_2,\Gamma_1,\Gamma_2,A,\alpha_1,\alpha_2 \right\rbrace = \left\lbrace .5,.01,.1 \, Gyr,0 \, Gyr,\frac{\ln(2)}{10 \ Gyr},\frac{\ln(2)}{10 \ Gyr},.025,\frac{1}{100 \, Gly^3 Gyr},\frac{100,000}{100 \, Gly^3 Gyr} \right\rbrace$.  (a) The default scenario is modified by the addition of a slower but more common competing strategy (subscript 2).  (b) Due to the common strategy, saturation is very rapidly visible from almost everywhere.  (c) Relative strategy success $A_{21}(t)$ changes by a factor of two during the transition as the saturation time delay becomes more significant in a crowded universe. (d) illustrates that first-mover advantage decays more rapidly for strategies with a large delay.}
\end{figure}

\subsection{Reheating}

In the default scenario, we illustrated a jump in the radiation energy-density of the universe as it became saturated with mass-consuming civilizations.  Here, we illustrate the change to the thermodynamic variables associated with the radiation component of the universe, with one caveat; we do not assume that such radiation is immediately in thermal equilibrium, only that it will thermalize eventually, and so we plot in figure 7 the quantities $\rho_{r}^{1/4}$ and $\rho_{r}^{3/4}$, which are always well-defined and will be proportional to the temperature and entropy density of radiation, respectively, in the fully thermalized case.

Specifically, two default scenarios are plotted with modifications to $A$ (the fraction of accessible dust mass) and $\Gamma$ (the decay constant) given by $ \left\lbrace \Gamma,A \right\rbrace = \left\lbrace \frac{\ln(2)}{ Gyr},.05 \right\rbrace$ and $\left\lbrace \Gamma,A \right\rbrace = \left\lbrace \frac{\ln(2)}{10 \, Gyr},.01 \right\rbrace$, representing particularly aggressive and tame use of resources, respectively, with corresponding changes to the radiation content of the universe.  In the tame scenario, the jump in $\rho_{r}^{3/4}$ (entropy density) is a single order of magnitude, whereas the aggressive scenario results in a jump approaching three orders of magnitude.  The jumps are calculated by comparing against the same quantities obtained from the standard cosmology, represented by the dashed lines.  The backreaction on the scale factor and Hubble parameter is also plotted, with fractional changes of order $10^{-5}$ in the tame scenario and approaching $10^{-3}$ in the aggressive scenario.

While the jumps in the radiation component of entropy are large in some scenarios, (with corresponding effects on the temperature, equation of state parameter, and backreaction on the expansion history) in discussions of the total entropy budget of the universe, one may be tempted to conclude that such changes are unimportant since they are dwarfed by the supermassive black hole contribution to total entropy~\cite{egan2010}.  This will be true if one includes the horizon entropy of black holes but excludes the entropy of the cosmic event horizon.  If one includes the cosmic event horizon, one finds a slightly different story -- due to the backreaction on the expansion history, the area of the cosmic event horizon (as viewed from the present, $t=13.75$) will be slightly enlarged by a factor of order $10^{-3}$ in the aggressive scenario described here.  While this is a small fractional increase in total entropy, it nevertheless represents a change that exceeds all local contributions to the entropy budget of the universe, including that of supermassive black holes, due to the enormous entropy of the cosmic event horizon which dominates all other sources.

\begin{figure}
\centering
\subfloat[]{
  \includegraphics[width=0.45\linewidth]{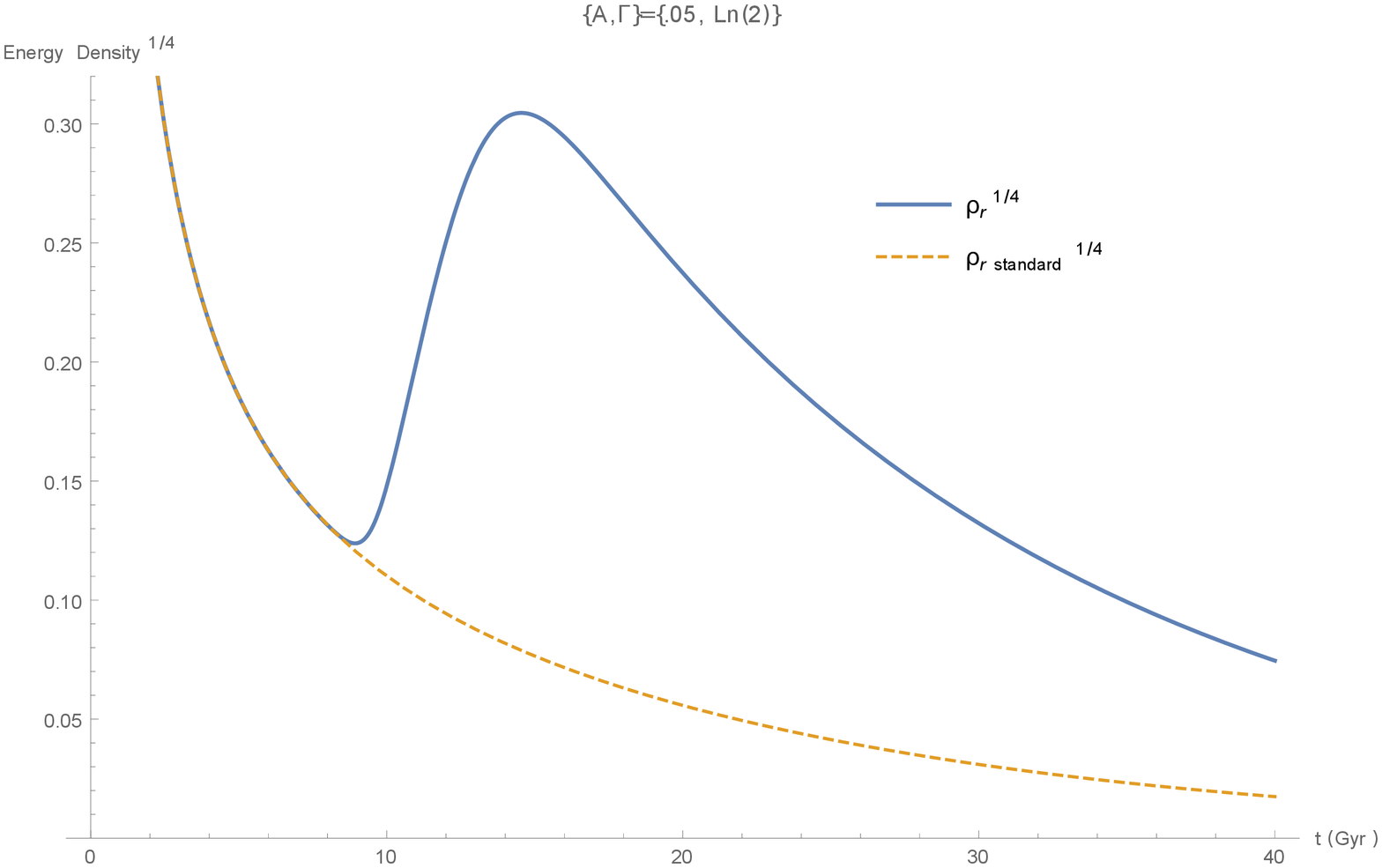}
}
\subfloat[]{
  \includegraphics[width=0.45\linewidth]{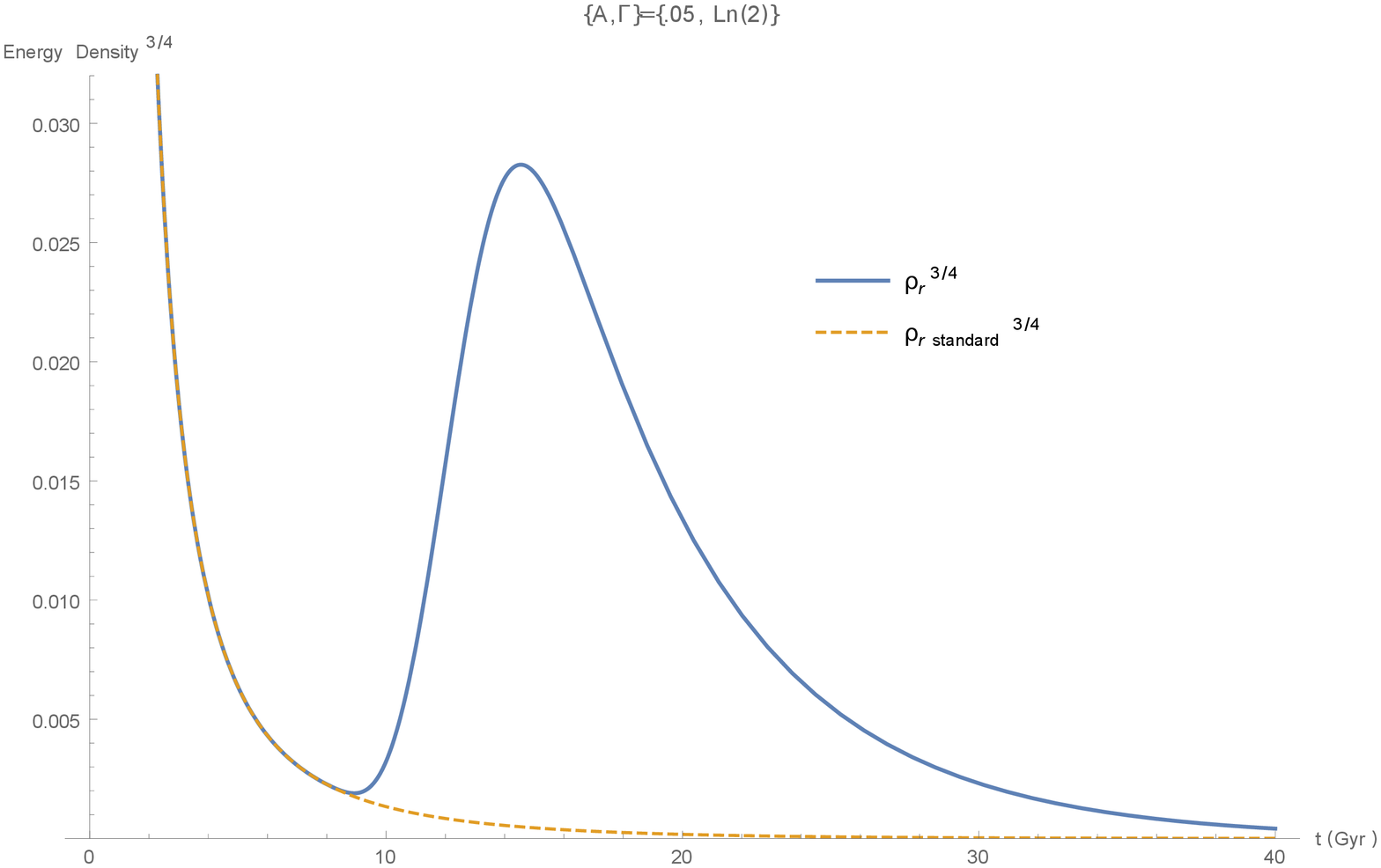}
}
\hspace{0mm}
\subfloat[]{
  \includegraphics[width=0.45\linewidth]{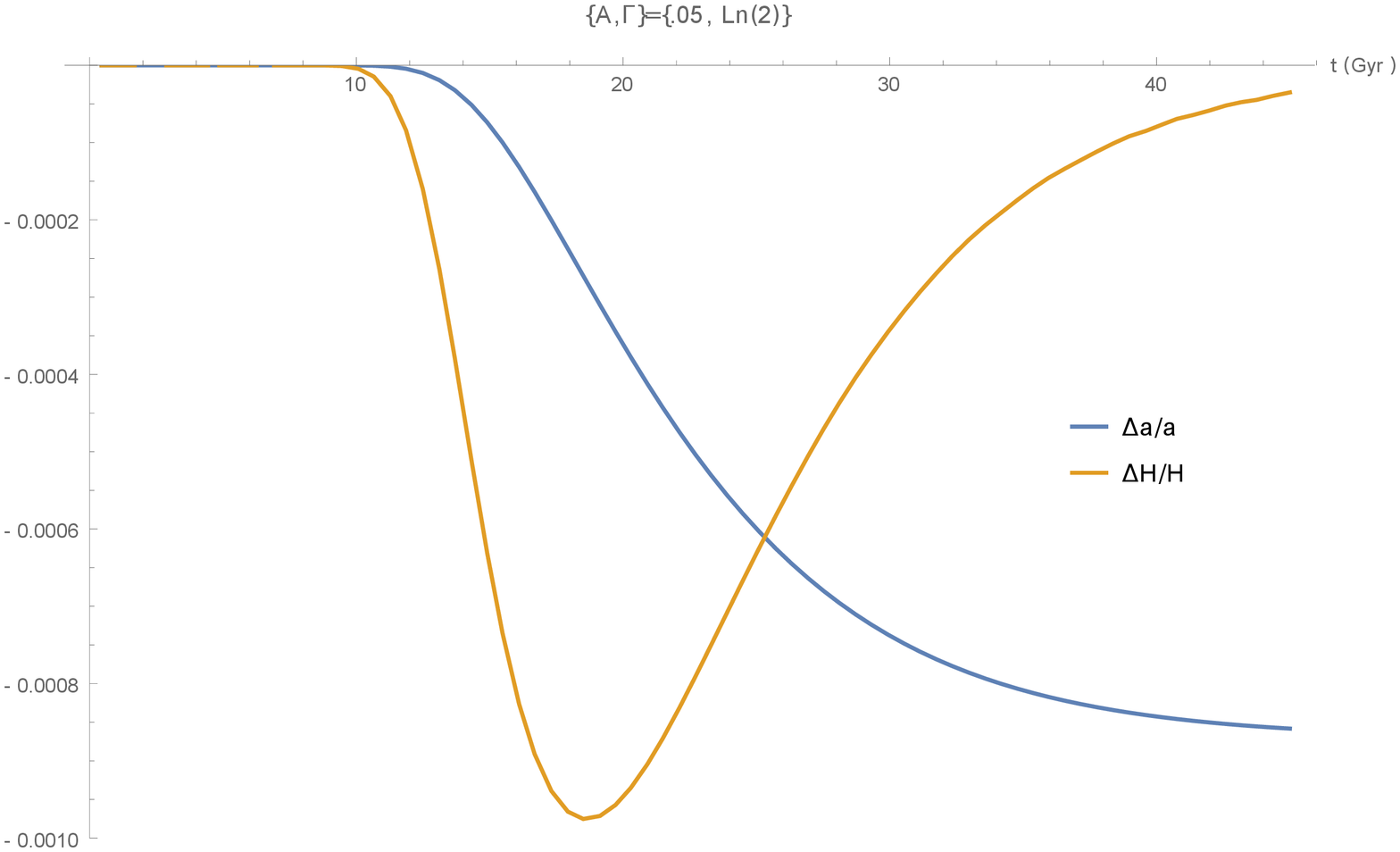}
}
\subfloat[]{
  \includegraphics[width=0.45\linewidth]{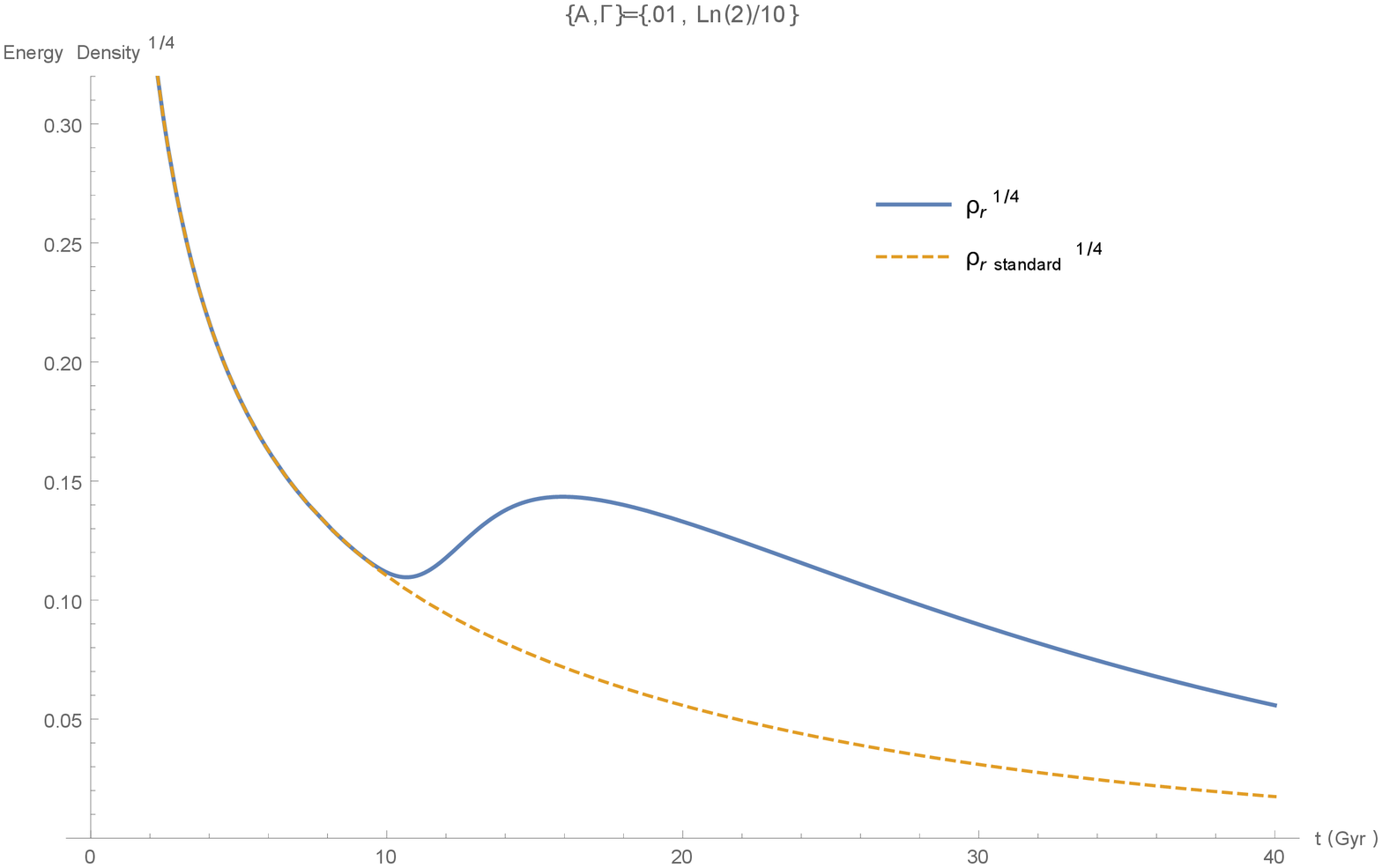}
}
\hspace{0mm}
\subfloat[]{
  \includegraphics[width=0.45\linewidth]{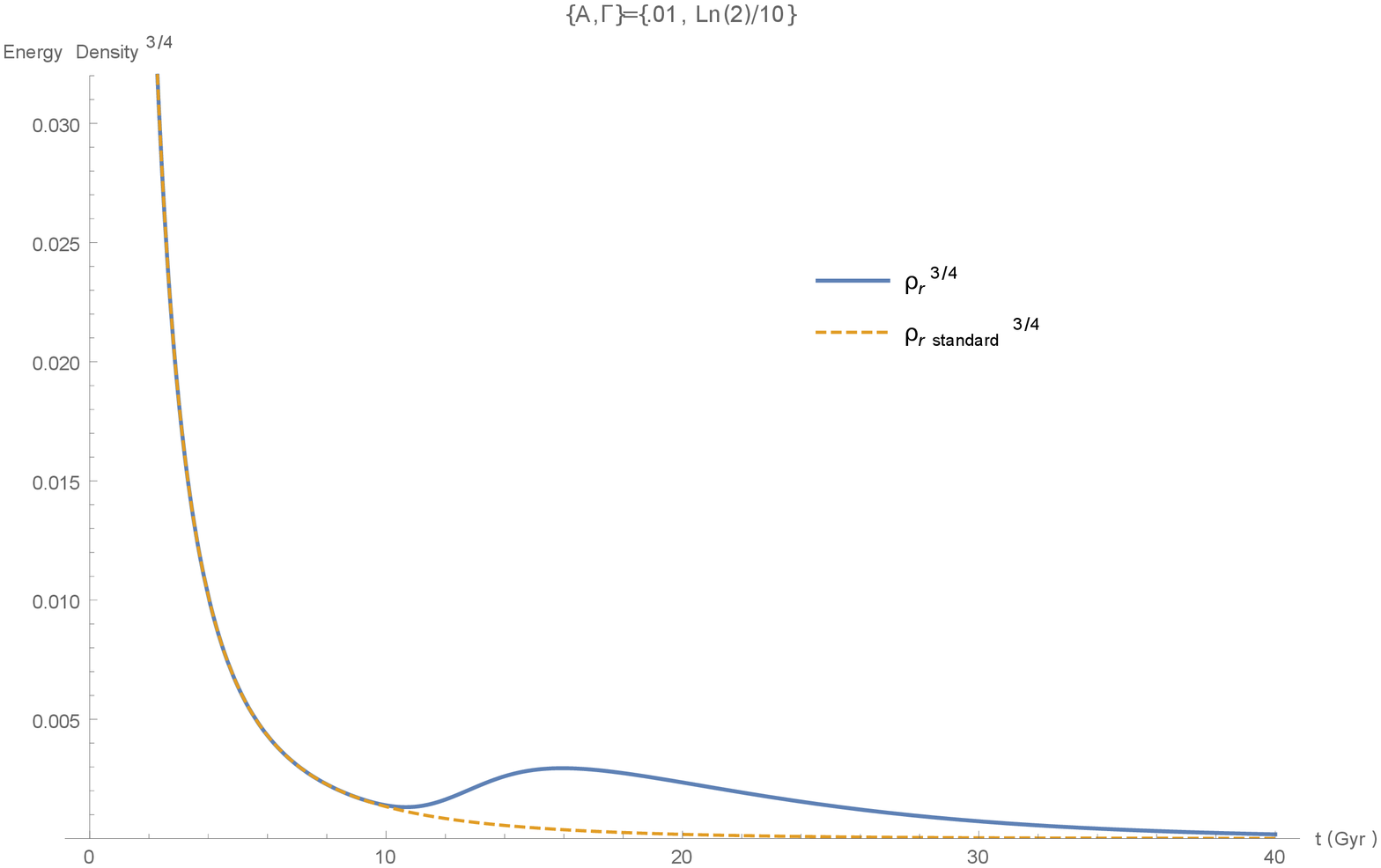}
}
\subfloat[]{
  \includegraphics[width=0.45\linewidth]{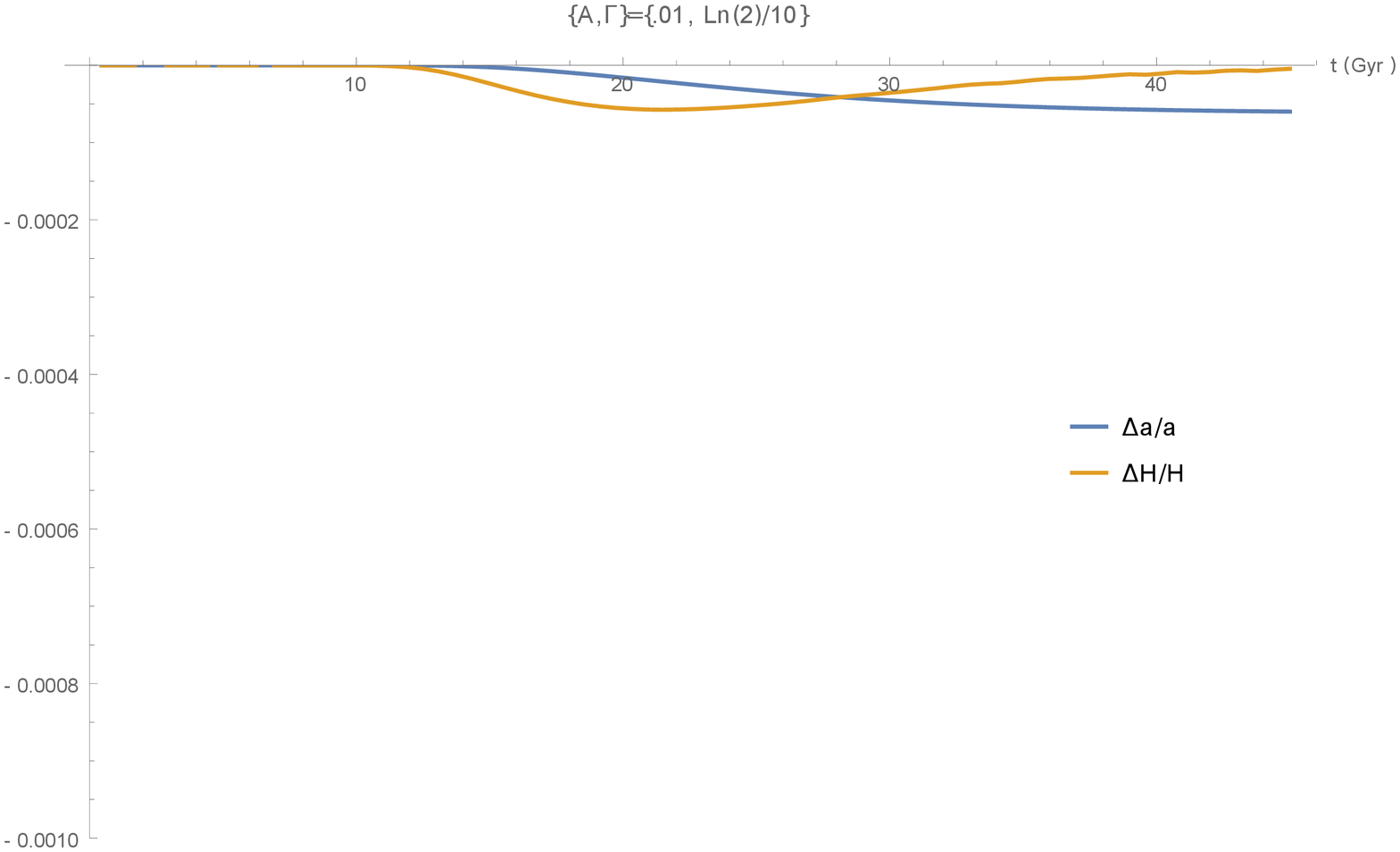}
}
\caption{Two reheating scenarios described by varying $ \left\lbrace \Gamma,A  \right\rbrace$ (rate of mass conversion and fraction of available mass) in the default scenario, representing aggressive and tame mass-consumption scenarios.  (a), (b), and (c) correspond to $ \left\lbrace \Gamma,A \right\rbrace = \left\lbrace \frac{\ln(2)}{ Gyr},.05 \right\rbrace$, while (d), (e), and (f) correspond to $\left\lbrace \frac{\ln(2)}{10 \, Gyr},.01 \right\rbrace$.  If the waste radiation were immediately thermalized, then $\rho_{r}^{1/4} \propto \mathrm{temperature}$ and $\rho_{r}^{3/4} \propto \mathrm{entropy \ density}$.  Note also the significant difference in the scale factor backreaction, between the two cases. }
\end{figure}

\subsection{Crossing scenarios}

Some strategies will strictly dominate others in terms of the final (average) volume controlled by the individual civilizations adopting them -- this will happen whenever both the probe front velocity $v$ is higher and the saturation time $T$ is lower than a competing strategy.  However, in the case where we compare a strategy with a higher $v$ to a strategy with a smaller $T$, the ranking of strategy success can become a time-dependent list (equivalently, relative strategy success $A_{i j}$ can transition from greater than to less than unity).  We illustrate this possibility in figure 8 with strategies that are modified from the default scenario by $ \left\lbrace v_1,v_2,T_1,T_2  \right\rbrace = \left\lbrace .5,.52,.01 \, Gyr,.1 \, Gyr  \right\rbrace$ -- i.e. strategy 1 is ``less delayed" whereas strategy 2 is ``faster."  If adopted sufficiently early, the faster-moving strategy 2 is expected to saturate more matter by the end of the transition, whereas adoption at late times (when crowding becomes a more important factor) will favor the less-delayed strategy 1.  Conversely, the scenario also shows that the less-delayed strategy 1 saturates more of the total universe early on, but is overtaken at late times by the faster-moving strategy 2.

\begin{figure}
\centering
\subfloat[]{
  \includegraphics[width=0.45\linewidth]{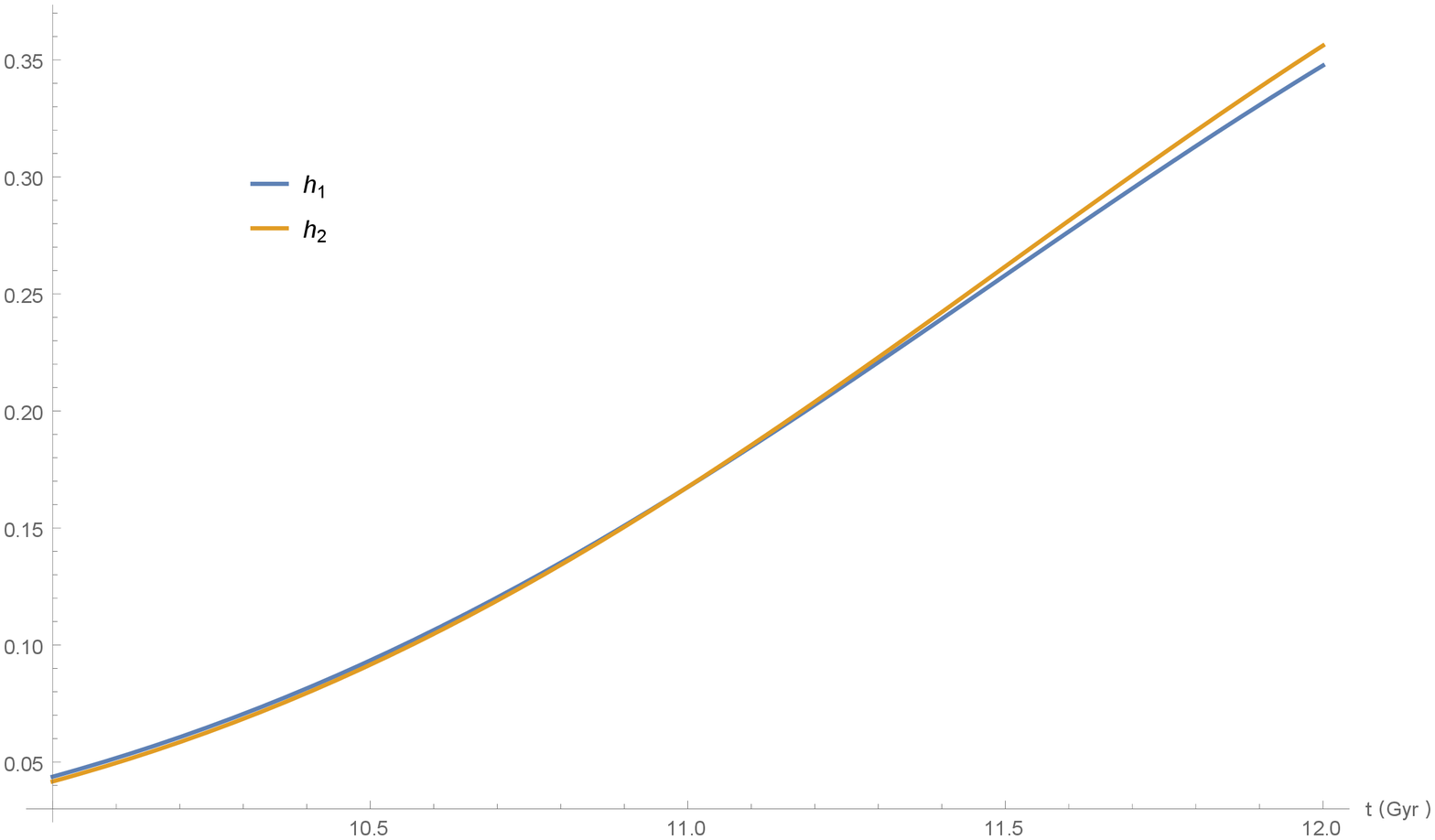}
}
\subfloat[]{
  \includegraphics[width=0.45\linewidth]{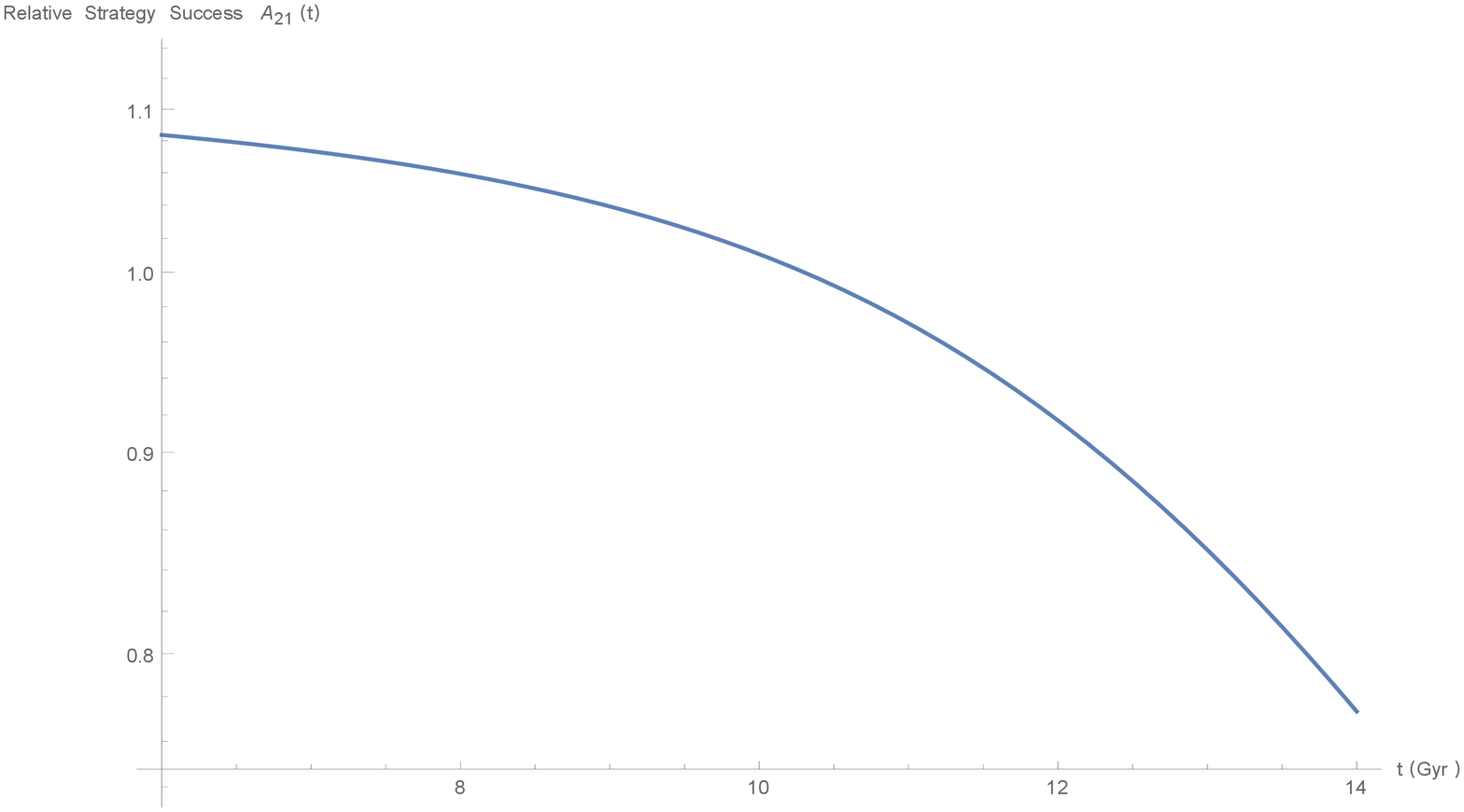}
}
\hspace{0mm}
\subfloat[]{
  \includegraphics[width=0.45\linewidth]{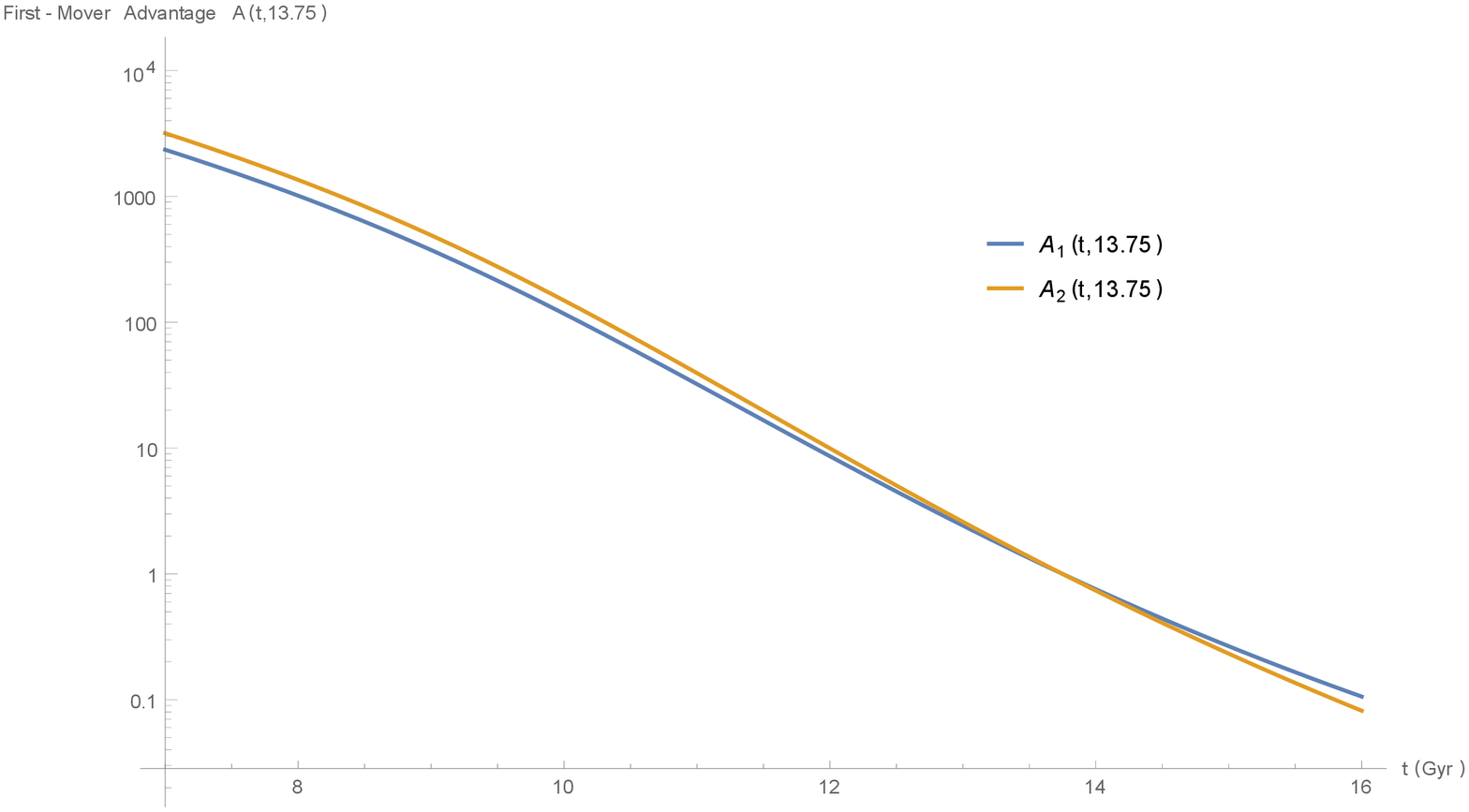}
}

\caption{A two-strategy scenario with multiple crossings, given by $ \left\lbrace v_1,v_2,T_1,T_2  \right\rbrace = \left\lbrace .5,.52,.01 \, Gyr,.1 \, Gyr  \right\rbrace$ (other parameters same as default).  (a)  The less delayed strategy 1 initially captures more of the universe, but is overtaken by the faster strategy 2.  (b)  From the point of view of an individual species, the faster strategy 2 saturates more space if the species appears sufficiently early -- appearing at later times, the less-delayed strategy 1 is more successful.  (c) The greater delay of strategy 2 results in a more rapid decay of first-mover advantage.
   }
\end{figure}

\end{widetext}
\section{Conclusions}

A feature common to every model we have explored is the very low appearance rate for aggressively expanding life.  This was done to obtain solutions in which the universe is not already completely saturated with life.  If one takes this as a \emph{requirement}, such that the apparent emptiness of our galaxy should not be an extremely improbable event, but simultaneously allow that technology will enable fast expansion speeds, then extremely low appearance rates become almost unavoidable.  This will mean that the characteristic distance to the nearest aggressive expansion event is likely to be very large from the viewpoint of life (such as humanity) that has found itself in an empty region of the cosmos.  We thus expect on quite general grounds that specialized observation techniques would be necessary to detect aggressive expansion events, rather than surveys of nearby galaxies within or close to our own supercluster.
 
Our most general conclusion is simply that the nature of life and the possible limits of technology, when taken together, should force us to reconsider the future large-scale evolution of the universe.  Life seeks out sources of thermodynamic free energy, and technology is the great enabler.  Having identified processes (such as black hole evaporation) which convert ``dead" mass directly into thermal radiation, one is immediately confronted with the question of how successful a fully life-saturated and technology-maximized universe will eventually be at exploiting them, and what the physical consequences will ultimately be.   

We have argued that the dynamics of nucleation and bubble growth in a first-order phase transition is naturally carried over to a description of aggressively expanding life saturating the universe, due to the geometrical similarities (spatially random events and spherical expansion).  However, due to the abundance of waste radiation appearing in this picture, it is tempting to regard the effects of aggressively expanding life as a \emph{literal} thermodynamic phase transition -- an abrupt change to the equation of state and thermodynamic variables describing the universe.  We are accustomed to regarding the thermal effects of life as due to very special initial conditions.  A Petri dish, for example, filled with nutrients and a few bacteria can be expected to have a very different thermodynamic future than a Petri dish filled with nutrients alone, due to the special initial conditions implied by specifying that bacteria are present -- the initial conditions are pre-configured for an immediate collapse in the free energy.  The cosmological process we have described, however, begins with no life initially, so it must represent a general kind of transition between states, i.e. one that is not merely a consequence of fine-tuning the initial conditions.  The statistical process by which the universe finds and abruptly transitions to the radiation-filled, higher-entropy state we have described is not through thermal fluctuations or tunneling directly to a new vacuum -- it is through the elaborate and indirect route of evolving life and general intelligence that forever hungers for new sources of free energy.

\section{Acknowledgements}

I am especially grateful to Bernard Yurke and Jonathan Olson for many stimulating conversations on the possible implications of life in the cosmos.

\section*{References}
\bibliographystyle{unsrt}
\bibliography{ref4}

\end{document}